\documentclass[preprint2]{aastex}

\usepackage{graphicx,epstopdf,amsmath}
\usepackage{import}

\usepackage{color}

\def\BB{{\bf {B}}}

\newcommand{\newa}{NewA}
\newcommand{\jkas}{JKAS}

\newcommand{\phpl}{PhPl}
\newcommand{\sci}{Sci}
\newcommand{\jcoph}{JCoPh}
\newcommand{\cophc}{CoPhC}

\newcommand{\soph}{SoPh}

\shorttitle{Simulated Solar Jets in Coronal Loops}
\shortauthors{Wyper \& DeVore}

\begin{document}

\title{Simulations of Solar Jets Confined by Coronal Loops}

\author{P.~F.~Wyper} 
\affil{Oak Ridge Associated Universities, Heliophysics Science Division, NASA Goddard Space Flight Center, 8800 Greenbelt Rd, Greenbelt, MD 20771}
\email{peter.f.wyper@nasa.gov}

\author{C.~R.~DeVore} 
\affil{Heliophysics Science Division, NASA Goddard Space Flight Center, 8800 Greenbelt Rd, Greenbelt, MD 20771}
\email{c.richard.devore@nasa.gov}

\begin{abstract}
Coronal jets are collimated, dynamic events that occur over a broad range of spatial scales in the solar corona. In the open magnetic field of coronal holes, jets form quasi-radial spires that can extend far out into the heliosphere, while in closed-field regions the jet outflows are confined to the corona. We explore the application of the embedded-bipole model to jets occurring in closed coronal loops. In this model, magnetic free energy is injected slowly by footpoint motions that introduce twist within the closed dome of the jet source region, and is released rapidly by the onset of an ideal kink-like instability. Two length scales characterize the system: the width ($N$) of the jet source region and the footpoint separation ($L$) of the coronal loop that envelops the jet source. We find that the jet characteristics are highly sensitive to the ratio $L/N$, in both the conditions for initiation and the subsequent dynamics. The longest-lasting and most energetic jets occur along long coronal loops with large $L/N$ ratios, and share many features of open-field jets, while smaller $L/N$ ratios produce shorter-duration, less energetic jets that are affected by reflections from the far-loop footpoint. We quantify the transition between these behaviours and show that our model replicates key qualitative and quantitative aspects of both quiet-Sun and active-region loop jets. We also find that the reconnection between the closed dome and surrounding coronal loop is very extensive: the cumulative reconnected flux at least matches the total flux beneath the dome for small $L/N$, and is more than double that value for large $L/N$.
\end{abstract}


\section{Introduction}
\label{sec:intro}

Observations of the Sun's outer atmosphere -- the chromosphere, transition region, and corona -- reveal the ubiquitous occurrence of jetting phenomena across a wide range of spatial and temporal scales. Transient, impulsive, collimated flows of plasma are observed both as bright, emitting features at high (coronal) temperatures and as dark, absorbing features at low (chromospheric) temperatures; generally, these events are referred to as jets and surges, respectively \citep[e.g.][]{Canfield1996}. The wavelengths involved range from the optical, through the UV and EUV, to X-rays. H$\alpha$ surges were identified and studied first, from ground observatories \citep[e.g.][]{Roy1973}. The advent of telescopes placed in space, above Earth's atmosphere, was a prerequisite for the detection of UV and EUV jets from {\it Skylab} \citep[][]{Schmahl1981} and from sounding rockets \citep[][]{Brueckner1983}, and, later still, X-ray jets from {\it Yohkoh} \citep[][]{Shibata1992,Shibata1994,Shimojo1996}. Cool surges and hot jets sometimes are observed together, in close association in space and time \citep[][]{Canfield1996}. Subsequent improvements in the spatial resolution and temporal cadence of space-borne instruments have enabled ever more detailed studies of jets and surges from the {\it Solar and Heliospheric Observatory} ({\it SOHO}) \citep[e.g.][]{Wang1998}, {\it Transition Region and Coronal Explorer} ({\it TRACE}) \citep[e.g.][]{Chae1999}, {\it Solar-Terrestrial Relations Observatory} ({\it STEREO}) \citep[e.g.][]{Patsourakos2008}, {\it Hinode} \citep[e.g.][]{Cirtain2007,Savcheva2007,Nishizuka2008,Torok2009,Moore2010,Moore2013,Liu2011}, {\it Solar Dynamics Observatory} ({\it SDO}) \citep[e.g.][]{Guo2013,Lee2013,Schmieder2013,Zheng2013,Shen2011}, and {\it Interface Region Imaging Spectrograph} ({\it IRIS}) \citep[e.g.][]{Tian2014,Cheung2015}.

Typically, an EUV or X-ray jet begins with a rapid brightening low in the solar atmosphere, indicating an impulsive increase in the plasma temperature or density, or both. This is followed by bulk outflows of material from the bright region, usually at supersonic speeds and highly collimated in direction. These properties suggest that the plasma flow is guided along the local magnetic field -- tracing loops in magnetically closed active regions and quasi-radial spires in magnetically open coronal holes -- and is subjected to nonthermal, presumably magnetic, forces. The bright emissions fade away gradually as the plasma cools and expands on its passage through the atmosphere and as the energy source driving the jet is depleted. In many jets, the outflows exhibit a distinctly helical structure in the loop or spire. Within this subclass, many events also display translational bodily motions of the jet loop or spire across the plane of the sky. With or without these helical and translational motions, a minority of jets are observed to recur from a single structure, yielding multiple episodes of brightening, jetting, and dimming that are separated by intervals of quiet.

\begin{figure}[t]
\centering
\includegraphics[width=0.45\textwidth]{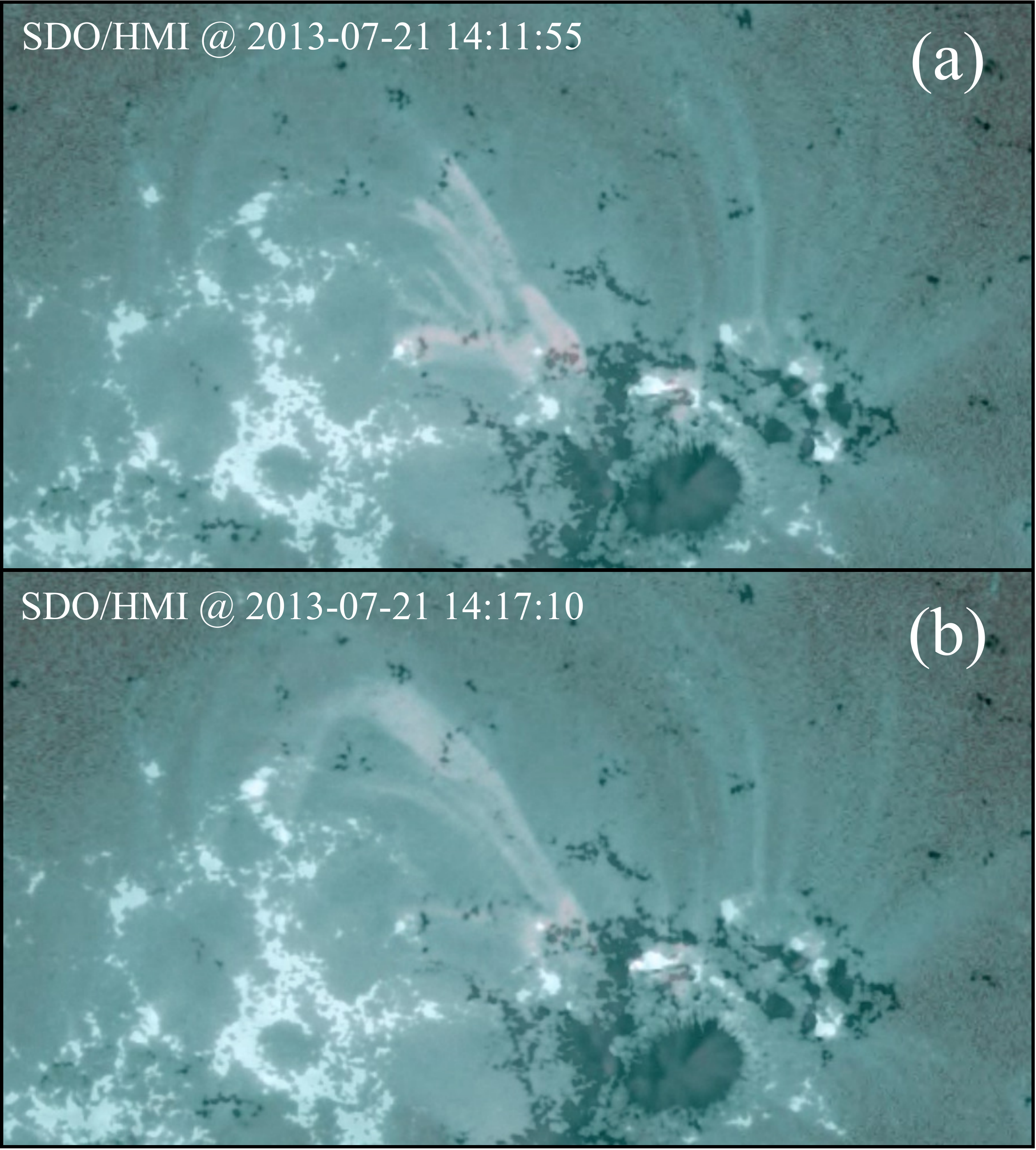}
\caption{A closed-loop jet observed by {\it SDO}, described in detail by \citet{Cheung2015}. HMI line-of-sight magnetograms (grayscale) overlaid with AIA 94\,\AA \ and 131\,\AA \ channel images (blue) show the (a) launch and (b) propagation to the far footpoint of a helical coronal-loop jet. An animation of this figure is available online.}
\label{fig:sdoaia}
\end{figure}

All of the above properties have been established by numerous reported observations of coronal-hole jets, which frequently are referred to as polar jets. These characteristics also have been observed, albeit for a much smaller event sample, in jets occurring in closed coronal loops \citep[e.g.][]{Torok2009,Yang2012,Guo2013,Lee2013,Schmieder2013,Zheng2013,Cheung2015}. The magnetic structures hosting these jets range in size from long loops rooted in weak-field areas adjacent to coronal holes, to short loops rooted in strong-field areas within young active regions. One example of a closed-loop jet, observed with {\it SDO}'s Atmospheric Imaging Assembly (AIA) and described in detail by \citet{Cheung2015}, is shown in Figure \ref{fig:sdoaia}. Our objective in this paper is to simulate jets in a variety of such closed coronal loops by employing a well-tested model, described below, that has been applied extensively to polar jets \citep{Pariat2009,Pariat2010,Dalmasse2012,Pariat2015}. 

\begin{figure}
\centering
\includegraphics[width=0.45\textwidth]{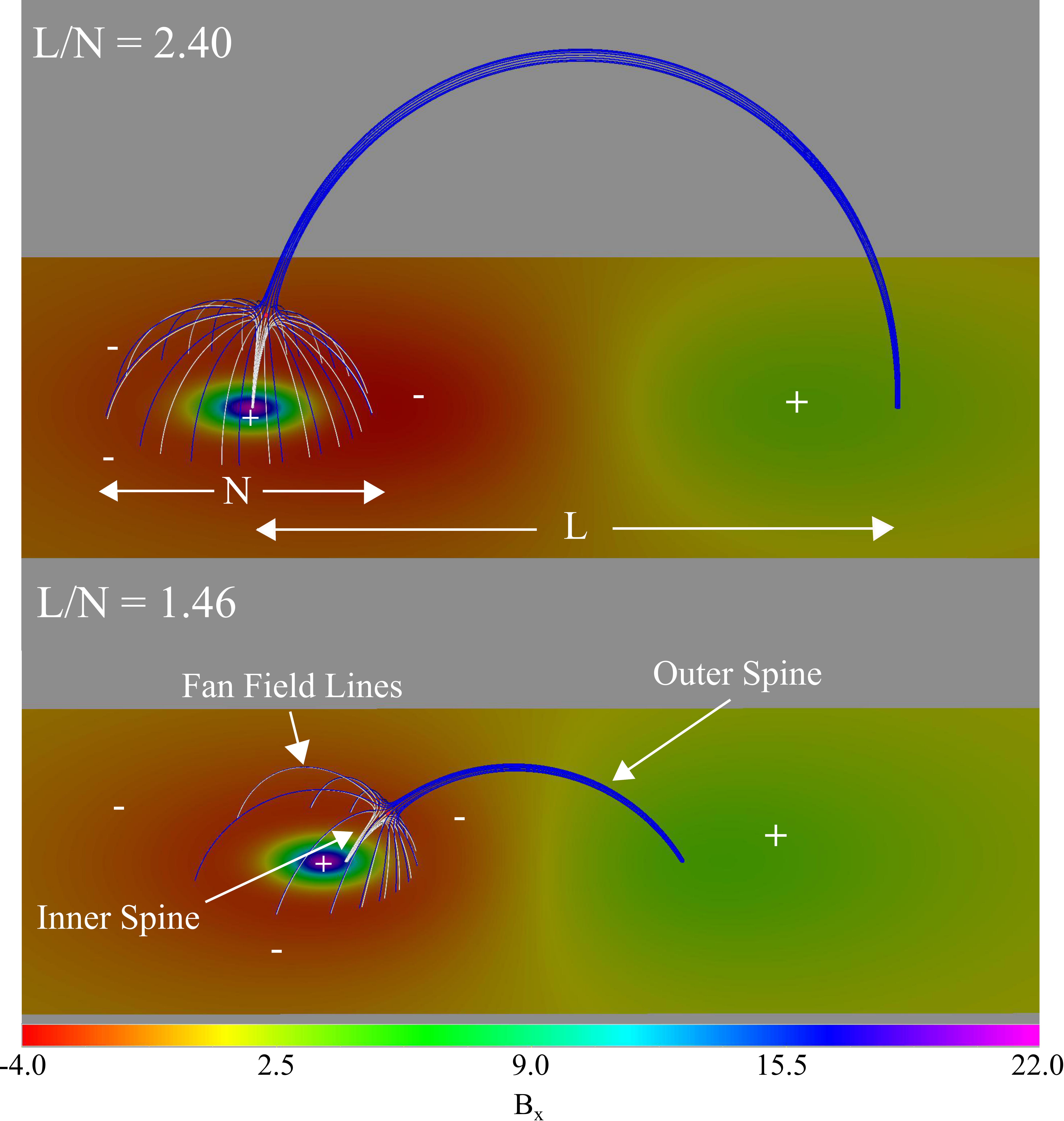}
\caption{Initial magnetic field in two configurations with aspect ratios $L/N=2.40$ (top) and $L/N=1.46$ (bottom). The bottom planes are color-shaded according to the sign ($+,-$) and strength of the field component normal to the surface ($B_x$). Selected magnetic field lines outline the fan separatrix surface and the inner and outer spine lines emanating from the null point. The diameter of the separatrix dome ($N$) and the separation of the spine-line footpoints ($L$) characterize the intrinsic spatial scales of the configuration.}
\label{fig:fields}
\end{figure}

The source region of many solar jets has the morphological appearance of a sea anemone \citep{Shibata1994}, with bright tendril-like curved loops emanating radially from a central locus. {It is well established from observations that such anemone regions form as a result of flux emergence into coronal holes \citep[e.g.][]{Torok2009,Liu2011}. The photospheric magnetic field beneath the loops typically has a concentrated patch of vertical field of one (minority) polarity embedded within a large-scale area of generally weaker vertical field of the other (majority) polarity \citep[e.g.][]{Cheung2015}. Magnetic flux tubes that originate within the minority-polarity patch close back to the Sun's surface locally, on the far side of the polarity inversion line encircling the patch. The manner in which the two polarity regions are connected depends on the free energy stored in this closed field. The minimum-energy potential magnetic field above such a distribution of photospheric flux has a dome-shaped structure with a magnetic null point near its top. Flux tubes inside of the dome form closed, unsheared loops like those observed in the anemone regions; flux tubes outside of the dome follow the large-scale background field associated with the majority polarity, closing back to the Sun farther away (in the case of closed-loop jets, as in the observational papers cited earlier and as shown in Fig.\ \ref{fig:sdoaia}) or opening out into the remote heliosphere (in the case of coronal-hole jets). Two of our model setups exhibiting a potential null dome embedded within closed coronal loops are shown in Figure \ref{fig:fields}. The separation of the magnetic configuration into distinct flux systems, one closed locally and the other open (or closed remotely, as in Fig.  \ref{fig:fields}), readily allows relative displacements of field lines to occur across the null point. Such displacements generate strong electric currents and, eventually, initiate magnetic reconnection and associated activity \citep{Lau1990,Antiochos1996}.}

A prototypical model for solar jets, accounting for their impulsiveness and their helical motions by appealing to magnetic reconnection between a closed volume of twisted magnetic flux and the ambient open untwisted field, was put forth by \citet{Shibata1986}. The reconnection transfers magnetic twist from closed to open field lines, and the twist residing on the newly reconnected open field lines then propagates away from the interaction region at the Alfv\'en speed to form an ``untwisting'' jet. This basic picture has been the basis for numerous simulations of jets driven by flux emergence through the solar photosphere, in which a twisted flux rope rises buoyantly into a pre-existing coronal magnetic field, initiates reconnection, and launches a jet when the field orientations are favorable \citep{Yokoyama1995,Yokoyama1996,Miyagoshi2003,Miyagoshi2004,Archontis2005,Galsgaard2005,Moreno-Insertis2008,Gontikakis2009,Torok2009,Archontis2010,Jiang2012,Archontis2013,Moreno-Insertis2013,Takasao2013,Fang2014}. This process also can form the anemone structure itself, as the observed end state of the evolution following closed-loop \citep{Torok2009} and coronal-hole \citep{Liu2011} jets.

It is by no means clear that significant flux emergence always precedes solar jets, however. The fundamental susceptibility of the null-point configuration to reconnection can be exploited in other ways, as noted by \citet{Shibata1986}. In studies of polar jets in open fields, Pariat and collaborators \citep{Pariat2009,Pariat2010,Pariat2015,Dalmasse2012} initiated reconnection and jet onset by slowly twisting the closed flux beneath the anemone dome with applied footpoint motions. When the amount of twist reaches a critical threshold, which depends upon the inclination of the ambient field to the surface, the closed flux convulses due to onset of an ideal kink-like instability \citep{Rachmeler2010}. The kink itself releases very little energy, but it initiates impulsive reconnection between the twisted closed flux inside and the untwisted open flux outside of the dome. This reconnection rapidly and efficiently releases a large fraction of the stored magnetic free energy, launching a very strong helical jet. In addition, as the reconnected open field ``untwists'' it carries the reconnection site around the dome, generating a translational bodily motion of the resulting jet spire. The reconnection and the jet flows subside gradually as the stored magnetic energy is depleted. If, however, the footpoint motions persist, then the free energy builds up again until another cycle of reconnection, jetting, and relaxation occurs, yielding recurrent events. All of these features are consistent with observed EUV and X-ray jets.

In this paper, we explore the ramifications for the embedded-bipole model of jets when the ambient magnetic structure is a closed coronal loop, rather than an open coronal hole. As may be anticipated, the behavior is quite similar to the polar case for long loops; for short loops, on the other hand, there are both quantitative and qualitative differences in the dynamics. We describe our results in the rest of the paper, which is organized as follows. In \S \ref{sec:model}, we describe the physical and numerical model. \S \ref{sec:longshort} compares and contrasts two representative cases, a long coronal loop and a short one, while \S \ref{sec:survey} describes the broader results of a parametric survey over loop length. We relate our simulations qualitatively and quantitatively to jet observations in \S \ref{sec:obs0}. A summary of the highlights of our results is given in \S \ref{sec:discuss}, along with a discussion of their implications for future research.

\section{Model}
\label{sec:model}
We consider the simplest possible magnetic configuration for a jet-generating anemone region, in which a small-scale patch of relatively strong vertical field of one polarity is embedded in a large-scale region of weaker vertical field of the opposite polarity. This large-scale field, in turn, belongs to a bipolar flux distribution that forms a long coronal loop. Two particular realizations of this configuration are shown in Figure \ref{fig:fields}. The topology is of the fan/spine type, with a magnetic null point atop an approximately hemispherical dome of fan field lines. Emanating from the null are inner and outer spine lines that root in the parasitic polarity and the far polarity of the background field, respectively. Two intrinsic length scales of this system are the diameter ($N$) of the separatrix dome and the separation ($L$) of the two spine footpoints. Both of these lengths are easy to measure numerically, and they can be approximated from observational data as the width of the brightened anemone region and the footpoint separation of the large-scale coronal loop enclosing the anemone, respectively. The aspect ratio $L/N$ of the configuration quantifies the relative sizes of the coronal loop and the enclosed separatrix dome, ranging from nearly unity when the background loop is almost as compact as the dome itself to very large values when the background loop is far larger in scale. Open-field jets correspond to the limit $L/N \rightarrow \infty$, where the ``coronal loop'' extends from the Sun out to the remote heliosphere.

The scenarios constructed and simulated numerically for this paper were all performed in non-dimensional units. This is convenient, given the broad ranges of scales of jets and coronal loops on the Sun. It is allowed due to the homogeneity of the magnetohydrodynamics (MHD) model, which permits fundamental scales for length, time, and mass to be factored out of the set of variables and equations. All of the dimensionless numbers governing the physics of the system -- the acoustic and Alfv\'en Mach numbers, plasma beta, Reynolds number, Lundquist number, etc. -- are preserved under such rescalings of the fundamental length, time, and/or mass. Later in \S \ref{sec:obs2}, we will scale our non-dimensional results by assuming some typical values for solar parameters, thereby deriving predicted properties of the simulated jets for comparison with observations.

\begin{table*}[ht]
\centering 
\begin{tabular}{c c c c c c c c c c c c c} 
\hline\hline 
$\left\vert y_{v} \right\vert$ & 4.0 & 4.5 & 5.0 & 5.5 & 6.0 & 6.5 & 7.0 & 7.5 & 8.0 & 8.5 & 9.0 & 10.0 \\ [0.5ex] 
$N$ & 5.86 & 5.82 & 5.82 & 5.84 & 5.88 & 5.96 & 6.08 & 6.18 & 6.34 & 6.48 & 6.69 & 7.04 \\ 
$L$ & 6.02 & 7.32 & 8.50 & 9.68 & 10.82 & 11.96 & 13.06 & 14.16 & 15.20 & 16.26 & 17.31 & 19.24 \\ 
$L/N$ & 1.03 & 1.26 & 1.46 & 1.66 & 1.84 & 2.01 & 2.15 & 2.29 & 2.40 & 2.51 & 2.59 & 2.73 \\ 
$\Psi_{dome}$ & 52.1 & 51.3 & 50.9 & 51.0 & 51.5 & 52.2 & 53.2 & 54.3 & 55.5 & 56.9 & 58.3 & 61.2 \\ 
$E_{inj}^{tot}$ & 115.5 & 118.1 & 120.3 & 122.0 & 123.9 & 124.6 & 126.4 & 124.6 & 131.9 & 132.7 & 133.4 & 134.5 \\
$t_{trig}$ & 1180.0 & 1010.0 & 840.0 & 790.0 & 720.0 & 655.0 & 535.0 & 560.0 & 570.0 & 650.0 & 790.0 & -\\ 
$t_{jet}$ & 80.0 & 130.0 & 100.0 & 130.0 & 120.0 & 165.0 & 260.0 & 300.0 & 315.0 & 330.0 & 330.0 & - \\[1ex] 
\hline\hline 
\end{tabular}
\caption{Simulation parameters: Vertical dipole position ($\left\vert y_v \right\vert$), separatrix dome diameter ($N$), length of outer coronal loop ($L$), aspect ratio ($L/N$), positive magnetic flux under the separatrix dome ($\Psi_{dome}$), total injected energy ($E_{inj}^{tot}$), and measured jet trigger times and durations ($t_{trig}$ and $t_{jet}$, respectively).}
\label{table:param} 
\end{table*}

\subsection{Initial Conditions}
\label{sec:initial}
The magnetic configurations illustrated in Figure \ref{fig:fields} are minimum-energy, electric current-free (``potential'') fields. To construct them, we superposed two magnetic dipoles: one horizontally oriented to generate the large-scale background field; the other vertically oriented with a more compact flux distribution at the coronal base to generate the parasitic polarity. By changing the strength and position of the vertical dipole, we varied the location and relative size of the separatrix dome compared with the coronal loop that encloses the dome.

\begin{figure}
\centering
\includegraphics[width=0.45\textwidth]{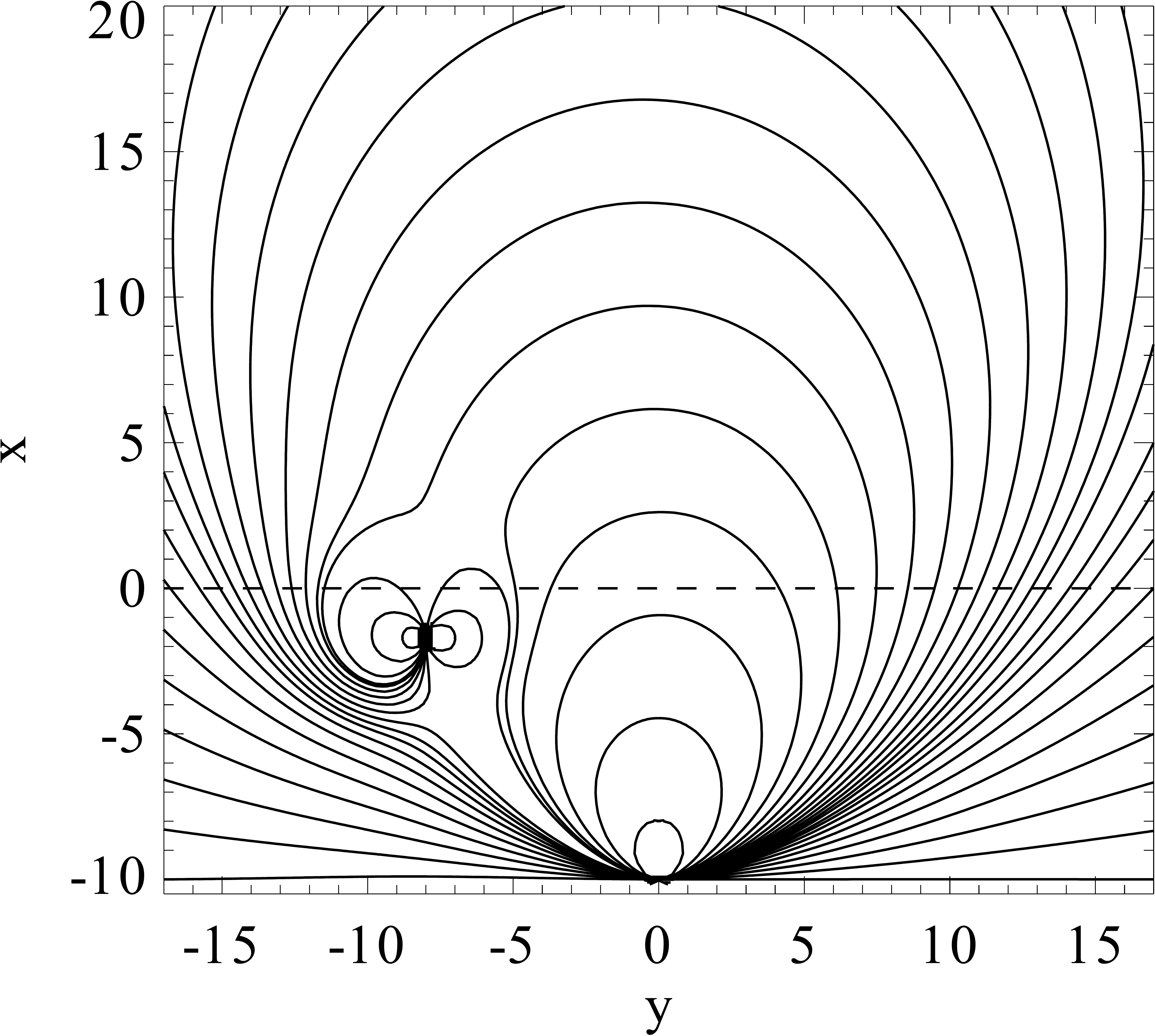}
\caption{Magnetic field lines (solid curves) in the $z=0$ plane show the configuration with $L/N=2.40$ constructed using two sub-photospheric magnetic dipoles. The dashed line shows the position of the photosphere, which is color-shaded in the top panel of Figure 2.}
\label{fig:dipoles}
\end{figure}

Specifically, the initial magnetic field is given by 
\begin{equation}
\mathbf{B} = \boldsymbol{\nabla} \times (\mathbf{A}_{h}+\mathbf{A}_{v}),
\end{equation}
where $\mathbf{A}_{h}$ and $\mathbf{A}_{v}$ are the vector potentials for the horizontal and vertical dipoles respectively. We orient the $x$ direction vertically in our simulation domain, with $x=0$ at the coronal base (the ``photosphere''), so the vector potentials can be written 
\begin{align}
\mathbf{A}_{h} &= \frac{B_{h}d_{h}^{3}}{2} \frac{(z-z_{h})\hat{\mathbf{x}}-(x-x_{h})\hat{\mathbf{z}}}{\left[(x-x_{h})^{2}+(y-y_{h})^{2} + (z-z_{h})^{2}\right]^{3/2}}, \nonumber\\
\mathbf{A}_{v} &= \frac{B_{v}d_{v}^{3}}{2} \frac{-(z-z_{v})\hat{\mathbf{y}}+(y-y_{v})\hat{\mathbf{z}}}{\left[(x-x_{v})^{2}+(y-y_{v})^{2} + (z-z_{v})^{2}\right]^{3/2}}.
\end{align}
We placed the background horizontal dipole below the photosphere at $(x_{h},y_{h},z_{h})=(-10.0,0.0,0.0)$, so its depth $d_h = 10.0$, with signed strength $B_{h} = +8.0$. This produces a peak field strength $\vert \BB_h \vert_{max} = 4.0$ at the photosphere above the horizontal dipole, $(x,y,z) = (0.0,0.0,0.0)$. The vertical dipole was placed below the photosphere at $(x_{v},y_{v},z_{v})=(-1.7,y_v,0)$, so its depth $d_v = 1.7$, with signed strength $B_{v} = +25.0$. We varied the $y$ position of the vertical dipole from $y_{v}=-4$ to $y_{v}=-10$. Its orientation is such that it opposes the direction of the background dipole field at its $y$ position, generating the parasitic polarity and creating the 3D null-point topology. Figure \ref{fig:dipoles} shows magnetic field lines in the $z=0$ plane and depicts the positions of the two magnetic dipoles relative to the photosphere for the case $L/N$ = 2.40. The peak vertical magnetic field above the parasitic polarity varied little ($B_x^{max} \approx 21$--$22$) over the range of $y_{v}$ values considered. The parameters of our various calculations are given in Table \ref{table:param}. These include $y_v$, $N$, $L$, $L/N$, and the positive magnetic flux under the separatrix dome, $\Psi_{dome}$. 

The peak magnetic pressure associated with our large-scale background field is $\vert \BB_h \vert_{max}^2/8\pi \approx 0.64$. We chose $P=0.01$ for the dimensionless thermal pressure, yielding a minimum plasma beta (ratio of thermal to magnetic pressure) of $\beta_h \approx 1.5 \times 10^{-2}$ with respect to the background field. At the center of the parasitic polarity, where the field strength is higher, the plasma beta is lower still, $\beta_v \approx 6 \times 10^{-4}$. The spatially varying field strength means that near the photosphere $\beta$ generally is well below unity and the magnetic field strongly dominates the dynamics, whereas high in our gravity-free corona, where the field is weaker but the thermal pressure remains uniform, $\beta$ reaches and then slightly exceeds unity. At and near the null point, in contrast, $\beta$ becomes very large (infinite, in principle, at the null) and the plasma dominates the dynamics, as it should. The majority of the jets in our parametric survey remain confined within the low-$\beta$ region; however, the experiments with $\left\vert y_{v} \right\vert \ge 8.5$ produce jets that impinge on the moderate-$\beta$ high corona. In these jets, the outflows reaching the apex of the loop bend the field lines rather sharply in the moderate-beta region. This bending of the field does not impede the progress of the jet, however, and the moderate $\beta$ in this region has no effect on the jet trigger and energy release, which occur well down in the low-$\beta$ portion of the corona.

We can set the mass density independently of the thermal pressure, so we chose an initial uniform value $\rho=1$.  The selection $R=0.01$ for the ideal gas constant then sets the units of temperature, with an initial uniform value $T=1$. These additional choices result in a uniform sound speed $v_{s}\approx 0.13$ throughout the domain, for a ratio of specific heats $\gamma = 5/3$. The same parameter values were assumed by \citet{Pariat2009,Pariat2010,Pariat2015}. The maximum Alfv\'{e}n speed is $v_{a} \approx 5.9$ at the center of the parasitic polarity. At the mean value across our sample of the aspect ratio, $L/N \approx 1.88$, the diameter of the closed separatrix dome is $N \approx 5.9$. Consequently, one dimensionless time unit in our simulations equals one characteristic Alfv\'{e}n time defined by these values for $v_{a}$ and $N$.

\subsection{Temporal Evolution}
\label{sec:evolve}
We use the Adaptively Refined Magnetohydrodynamics Solver \citep[ARMS;][]{DeVore2008} to solve the ideal MHD equations in the form 
\begin{gather}
\frac{\partial \rho}{\partial t} + \boldsymbol{\nabla}\cdot(\rho \mathbf{v}) = 0, \\
\frac{\partial (\rho \mathbf{v})}{\partial t}+\boldsymbol{\nabla}\cdot(\rho \mathbf{v}\mathbf{v}) + \boldsymbol{\nabla} P -\frac{(\boldsymbol{\nabla}\times \mathbf{B})\times\mathbf{B}}{\mu_{0}} = 0, \\
\frac{\partial U}{\partial t}+\boldsymbol{\nabla}\cdot (U\mathbf{v})+P\boldsymbol{\nabla}\cdot\mathbf{v} = 0,\\
\frac{\partial \mathbf{B}}{\partial t}-\boldsymbol{\nabla}\times (\mathbf{v}\times\mathbf{B})=0.
\end{gather}
Here $t$ is the time, $\rho$ the mass density, $P = \rho R T$ the thermal pressure, $U = P/(\gamma-1)$ the internal energy density, $\mu_{0}=4\pi$ the magnetic permeability, and $\mathbf{B}$ and $\mathbf{v}$ the 3D magnetic and velocity fields. Reconnection occurs in our simulations through numerical diffusion terms in the Flux-Corrected Transport scheme \citep{DeVore1991} employed by ARMS. The lack of an explicit resistivity allows the greatest amount of free magnetic energy for the given resolution to be built up in our calculations before rapid reconnection is initiated and the jet occurs. The FCT solution algorithm adds explicit numerical diffusion to the equations, and a corresponding amount of explicit anti-diffusion that minimizes the residual truncation error in smooth regions of the flow. These error-cancelling anti-diffusion fluxes are limited (`corrected'), however, so that artificial numerical ripples are suppressed at shock fronts, shear layers, current sheets, and other discontinuities, as described in detail by \citet{DeVore1991}. The effect is to introduce just enough dissipation to keep the solution well-behaved when ideal motions, such as the onset of kink instability in the case of our jet simulations, induce sudden, strong changes in the magnetic field and the plasma variables.

We energize the system in the same manner as \citet{Pariat2009}. That is, we impose a slow, subsonic and sub-Alfv\'{e}nic, rotation of the parasitic polarity that introduces magnetic shear across the nearly circular polarity inversion line. Specifically, the driving profile is
\begin{align}
\mathbf{v}_{\perp} &= v_{0}f(t)g(B_x)\hat{\mathbf{x}}\times \boldsymbol{\nabla} B_{x},\nonumber\\
f(t) &\equiv \frac{1}{2}\left[1-\cos\left(2\pi\frac{t}{t_{twist}}\right)\right],\nonumber\\
g(B_x) &\equiv k_{B}\frac{B_{r}-B_{l}}{B_{x}}\tanh\left(k_{B}\frac{B_{x}-B_{l}}{B_{r}-B_{l}}\right).
\label{eq:vdef}
\end{align}
The above expressions are used for $t \in [0,t_{twist}]$ and $B_{x} \in [B_{l},B_{r}]$; outside of those intervals, we set $\mathbf{v}_{\perp} = 0$. {The flow follows the contours of $B_{x}$ within the parasitic polarity patch and is incompressible, i.e. divergence-free, so that it preserves the vertical component of the magnetic field at the photosphere throughout the evolution.} The flow speed vanishes exactly as $B_{x} \rightarrow B_{l}$, and it becomes small as $B_{x} \rightarrow B_{r}$ because $\nabla B_{x}$ becomes small at the center of the parasitic polarity patch. We chose $t_{twist}=1000$, $B_{l}=0.6$, $k_{B}=5.0$, and $v_{0}=6.84\times10^{-5}$; $B_{r}$ was varied over the narrow range $[20,21]$, depending upon the peak strength of the dipole, to rotate the dipole as close to its central axis as possible. For these choices, between $0.7$ and $1.1$ maximum turns of twist, $M$, are injected into the field beneath the dome by the time of jet initiation, as shown below. The peak driving speed on the boundary is $\vert \mathbf{v}_{\perp} \vert \approx 0.016$, which is about $12\%$ of the sound speed and $0.3\%$ of the local Alfv\'{e}n speed. Thus, the magnetic field evolves quasi-statically and remains approximately force-free throughout the low-$\beta$ portion of the corona. 

The adaptive grid employed by ARMS is constructed from a basis set of root blocks, which can be subdivided to attain higher grid refinements in a pre-defined way and/or adaptively as the solution requires \citep{MacNeice2000}. The root blocks in these simulations had a fixed spatial extent of $17\times17\times17$. For values of $\left\vert y_{v} \right\vert$ in the range $[4,8]$, a domain size of $[0,34] \times [-17,17] \times [-8.5,8.5]$ ($2\times2\times1$ blocks) was sufficient to avoid any significant influence of the boundaries during the jet evolution. For larger values of $\left\vert y_{v} \right\vert$, a larger domain was necessary. To maintain the same grid spacing in these calculations, the number of root blocks was increased to $2\times3\times2$, giving a domain size of $[0,34] \times [-25.5,25.5] \times [-17,17]$. In each calculation, we required a minimum grid refinement of $4$ and a maximum of $6$ from the initial $32 \times 32 \times 16$ grid (on the smaller domain). A volume in each simulation that encompasses the footprint of the separatrix dome, with a height of $\approx 0.5$, is fixed at the maximum refinement level to resolve the boundary driving motions as finely as possible.  The grid outside this volume adapts according to whether strong gradients, beyond a fixed dimensionless value, develop in the magnetic field \citep[for details, see][]{Karpen2012}. This adaptive refinement better resolves the electric current layers and any shocks that develop in the domain. For a typical calculation on the smaller domain this results in approximately $1.2\times10^6$ grid cells, compared with around $5.4\times10^8$ for the equivalent grid uniformly refined to the same maximum resolution everywhere. The larger domain required only slightly more grid cells, because the outer regions where the domain was extended were refined almost solely to the minimum level of $4$. All boundaries of the box were closed (zero fluxes of mass, momentum, energy, and magnetic flux passed through) and line-tied ($\mathbf{v} = 0$ except where nonzero $\mathbf{v}_{\perp}$ was imposed according to Equation \ref{eq:vdef}).

\section{Results I: Long and Short of Loop Jets}
\label{sec:longshort}
We present our results by first focusing on the differences between the jets produced in the two configurations shown in Figure \ref{fig:fields}: a short coronal loop with small aspect ratio, $L/N = 1.46$, and a long loop with large aspect ratio, $L/N = 2.40$. The latter value is large enough to exhibit dynamics that are markedly different from the former, while also resembling rather closely those of the open-field jets simulated by \citet{Pariat2009,Pariat2010,Pariat2015}. These two cases fall near the extremes of our parameter range, which will be explored more fully below in $\S$\ref{sec:survey}. The variation in $L/N$ is determined principally by the coronal loop length, $L \in [6.0,19.2]$, whereas the size of the separatrix dome is roughly fixed, $N \in [5.8,7.0]$; see Table \ref{table:param}.

There are two aspects to the dependence upon $L/N$. First, the explosive reconnection that gives rise to a jet is of finite duration, as is the time required for disturbances launched by the jet to reach the far footpoints of the coronal loop and reflect back into the jet source region. For long loops, the duration is less than the travel time, and the jet generation process essentially is unaffected by the fact that the enveloping structure closes back to the Sun. For short loops, on the other hand, the duration can be longer than the travel time, and reflected disturbances can impact the jet source region before the generation process is complete. Our two examples illustrate this distinction. Second, the position of the null point on the separatrix dome is sensitive to the inclination of the background horizontal dipole field above the position of the embedded polarity. For long loops, the orientation is nearly vertical, the null point is positioned near the top of the dome, and the entire dome is nearly axisymmetric. For short loops, in contrast, the orientation is far from the vertical, the null point is positioned well over on the side of the dome nearest the far loop footpoint, and the dome is strongly asymmetric. In the latter case, it is much easier to displace the inner and outer spine lines along the fan surface, thereby distorting the potential null point into a current patch where reconnection can occur \citep{Antiochos1996,Pontin2007}. In the former case, the reconnection is impeded very effectively until onset of an ideal kink-like instability strongly breaks the near-axisymmetry of the dome. A survey of simulated open-field jets with varying tilt angle of the uniform background field \citep{Pariat2015} confirms this expected range of behaviors. We anticipated that our more asymmetric, short-loop configuration would form a current layer more readily and that reconnection would play more of a role at all stages of the evolution than for our more symmetric, long-loop configuration. As detailed below, this is just what we observe.

In all of the open-jet calculations of \citet{Pariat2009,Pariat2010,Pariat2015}, the system evolved through three main phases. Our new closed-field simulations also exhibit this progression. During the initial, energy-storage phase, the twist slowly builds in the field beneath the separatrix dome and magnetic free energy accumulates in the structure. There is little to no energy release due to any slow reconnection that occurs at the strengthening null-point current patch. Eventually, a critical threshold for magnetic twist or free energy -- governed by the ideal kink-like instability for the axisymmetric case, at least -- is attained, the separatrix dome convulses, and the impulsive energy-release phase begins. Rapid spine-fan reconnection \citep{PriestPontin2009} is initiated across the separatrix, releasing much of this twist and free energy onto external, shear-free field. The untwisting of the newly reconnected field lines causes the reconnection site to precess around the separatrix dome, generating nonlinear torsional Alfv\'{e}n waves that comprise a helical jet \citep{Patsourakos2008,Pariat2009}. Subsequently, these waves propagate away from the separatrix dome during a concluding relaxation phase, transporting significant magnetic energy and helicity away from the jet source region and causing the jet to subside as the field relaxes toward a lower energy state.

\begin{figure*}[t]
\centering
\includegraphics[width=0.99\textwidth]{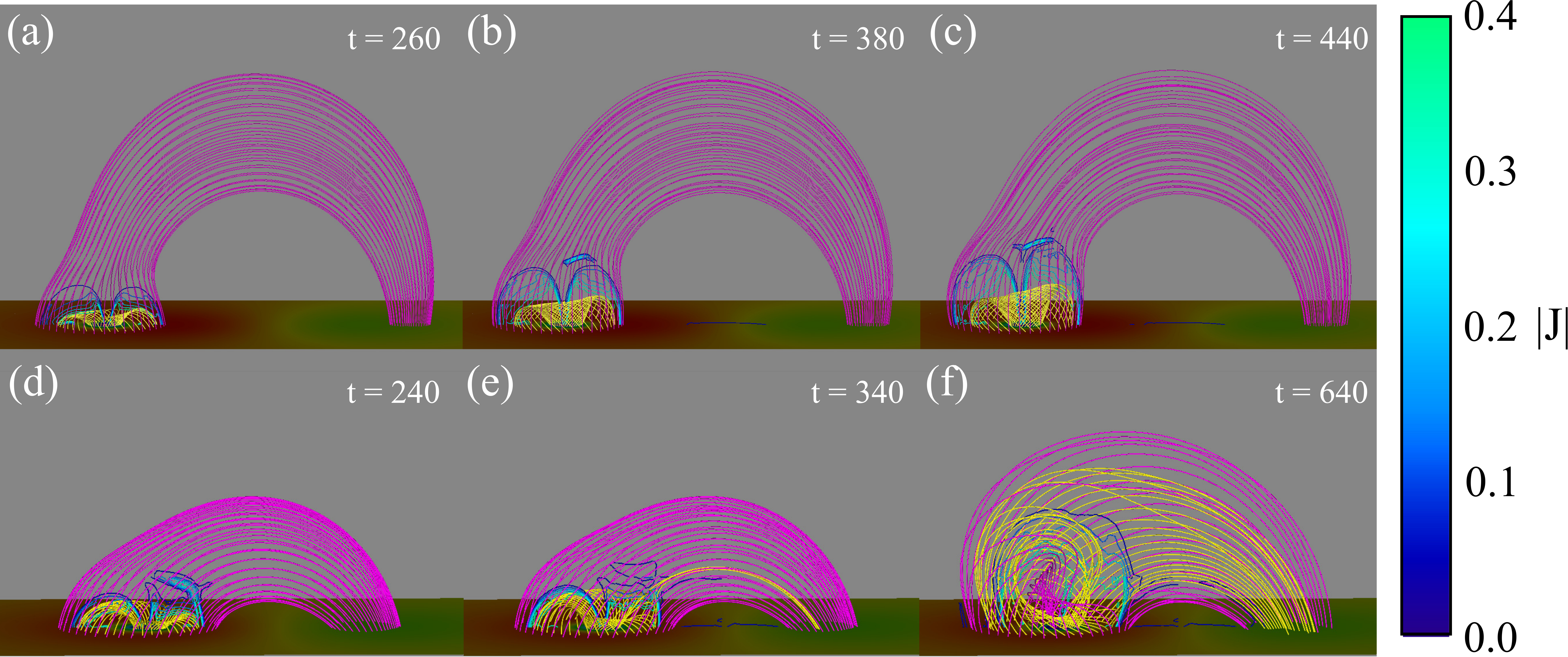}
\caption{Energy buildup phase for $L/N=2.40$ (top) and $1.46$ (bottom). Yellow and purple magnetic field lines are traced from fixed, line-tied footpoints that at $t=0$ reside inside and outside of the separatrix, respectively. The horizontal-plane shading shows $B_x$ as in Figure \ref{fig:fields}. The vertical-plane contours show the current-density magnitude $|\mathbf{J}|$, whose scale is saturated for clarity. Animations of both configurations are available online.}
\label{fig:buildup}
\end{figure*}

The following subsections qualitatively compare and contrast the three phases of jet evolution for our two selected cases.  A final subsection analyzes the rate and location of the reconnection that occurs in these events before a quantitative discussion of the full parameter study in $\S$\ref{sec:survey}.

\subsection{Energy Storage Phase}
\label{sec:pre-jet}
The evolution of the system as twist is introduced by the boundary driving motions prior to the jet is shown in Figure \ref{fig:buildup}. For the long-loop case ($L/N=2.40$, Fig.\ \ref{fig:buildup}a-c), the closed flux within the separatrix dome (yellow field lines) expands upward as the magnetic pressure increases beneath the dome. A current layer of small spatial extent gradually forms about the null as the dome expands into the surrounding loop flux (Fig.\ \ref{fig:buildup}b-c). The reconnection associated with this current layer is small, and the outflows are weak, so that very little of the flux beneath the dome reconnects prior to jet onset. In contrast, for the short-loop case ($L/N = 1.46$, Fig.\ \ref{fig:buildup}d-f), a much more extended current layer quickly forms around the null and the nearby separatrix surface as the dome expands. The resulting reconnection links flux previously closed beneath the dome to the far coronal loop footpoint (Fig.\ \ref{fig:buildup}f, yellow field lines), thereby transferring part of the injected twist onto the enclosing loop. Indeed, in \S \ref{sec:char} we will show that almost all of the flux beneath the dome is reconnected during this phase prior to onset of the short-loop jet.

\subsection{Energy Release Phase}
\label{sec:jet}
\begin{figure}[t]
\centering
\includegraphics[width=0.45\textwidth]{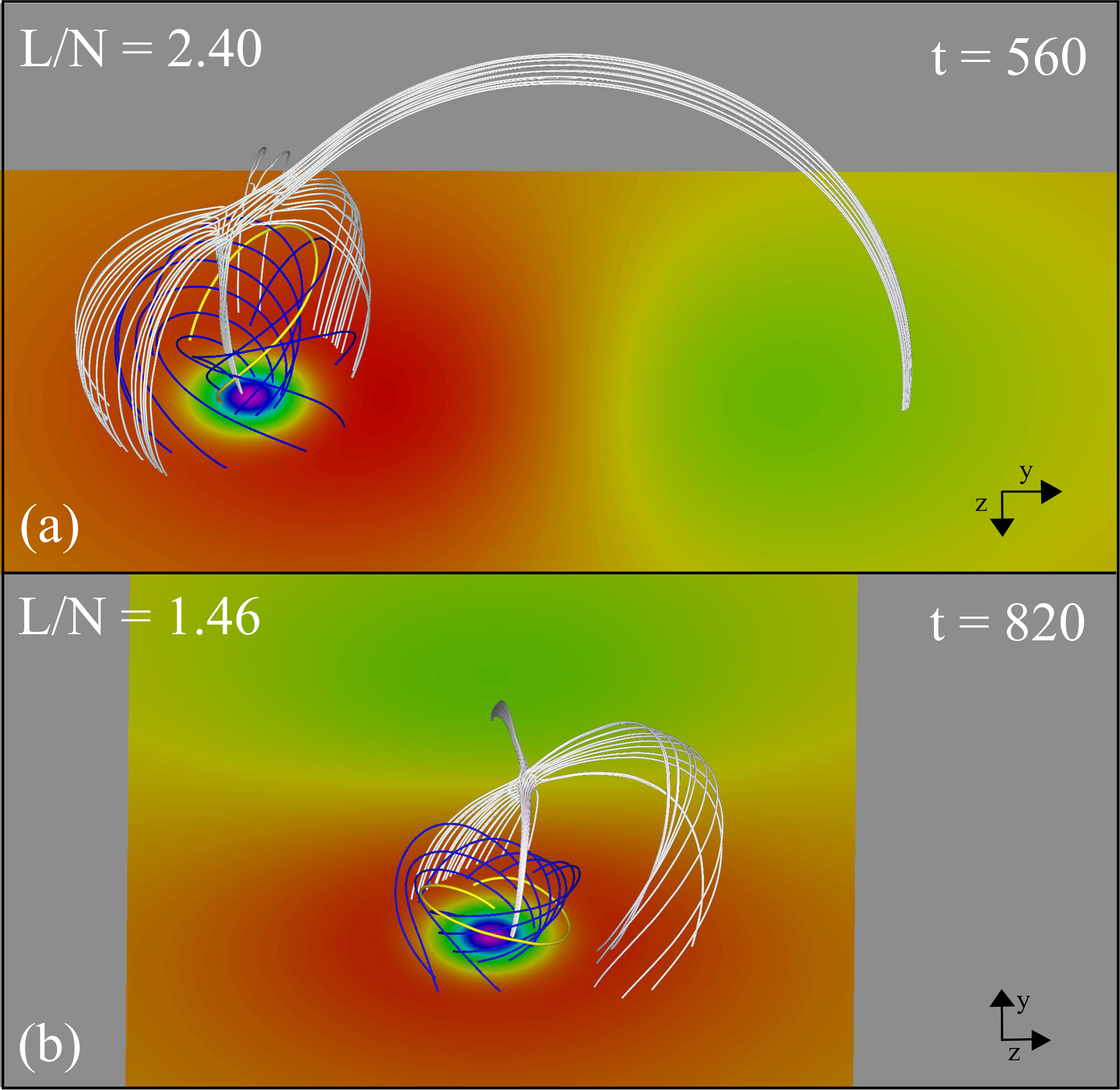}
\caption{Injected twist just prior to jet initiation. The magnetic field line with the highest number of turns about the inner spine is colored yellow. Blue field lines are traced from footpoints evenly spaced along the contour of $B_{x}$ (horizontal-plane shading) where the yellow field line starts. Silver field lines show the field structure near the null region.}
\label{fig:twist}
\end{figure}

In both cases, the initiation of fast energy release appears to be driven by the onset of a kink-like instability. Figure \ref{fig:twist} shows the field line (yellow) with the greatest number of turns beneath the dome prior to the initiation of the impulsive jet in each case ($M_{trig} \approx 0.8$ and $\approx 1.1$ turns for $L/N = 2.40$ and $1.46$, respectively). Also shown are field lines (blue) traced from evenly spaced footpoints around the contour of $B_{x}$ from which the yellow field line begins. Because our boundary driving follows the contours of $B_{x}$, albeit at non-constant speed, under ideal evolution the blue and yellow field lines should have approximately the same number of turns about the inner spine of the null in each simulation. For the long loop with $L/N=2.40$, this is indeed the case, showing that the weak reconnection near the null has had little effect on the most sheared magnetic flux. The approximate cylindrical symmetry of the most strongly sheared field is maintained up to the time of initiation of the jet. The sudden breaking of the symmetry and onset of impulsive reconnection closely resembles the evolution of previous open-field calculations with perfect initial symmetry \citep{Pariat2009,Rachmeler2010}, and is consistent with onset of a kink-like instability. The critical number of turns at onset, $M_{trig} \approx 0.8$, is very close to that reported by \citet{Pariat2010} when the background open field is tilted from the vertical by 10$^\circ$. Our long-loop configuration has an effective tilt angle of about 9$^\circ$ (see \S \ref{sec:initiate} and Fig.\ \ref{fig:theta} for more details).

For the short loop with $L/N=1.46$, in contrast to the long loop, the early reconnection reaches field lines that are close to the PIL, along most of its extent. This acts to reduce the shear of the field lines that straddle the PIL so that some field lines on the same contour of $B_{x}$ are more sheared than others. Along a given contour, the unreconnected field (having the greatest number of turns) folds underneath these less-twisted field lines (Fig.\ \ref{fig:twist}b, yellow field line), thereby contorting the shape of the separatrix. Thus, the early reconnection in the short-loop case reduces the amount of flux beneath the dome that is most strongly twisted prior to initiation of the instability. It also enhances the cylindrical asymmetry of the configuration. Nevertheless, the maximum number of turns attained ($M_{trig} \approx 1.1$) is well above that achieved ($\approx 0.8$) prior to the long-loop jet, while still well below that of the perfectly symmetric open case ($\approx 1.4$). This is also true of the average number of turns (\S \ref{sec:initiate} and Fig.\ \ref{fig:triggers}). Evidently, the critical twist for onset of the kink-like instability driving these impulsive jets is not a simple, monotone function of the effective tilt angle of the configuration.

\begin{figure*}[t]
\centering
\includegraphics[width=0.99\textwidth]{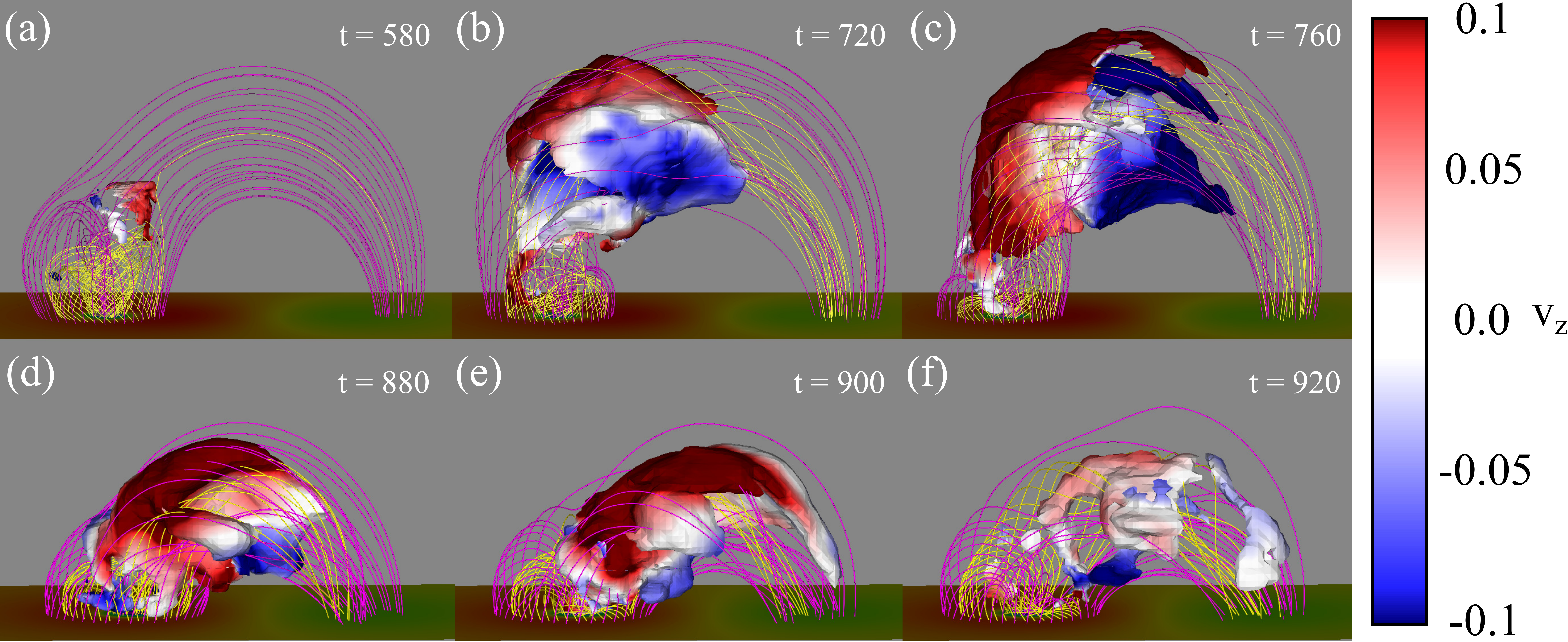}
\caption{The impulsive jet produced when $L/N = 2.40$ (top) and $1.46$ (bottom) with magnetic field lines as in Figure \ref{fig:buildup}. Isosurfaces of velocity magnitude $|\mathbf{v}| = 0.14$ are color-shaded according to $v_{z}$, the out-of-plane velocity component, whose scale is saturated for clarity. The horizontal-plane shading shows $B_x$ as in Figure \ref{fig:fields}. Animations of both configurations are available online.}
\label{fig:jets}
\end{figure*}

The jet produced in our long loop ($L/N=2.40$; Fig.\ \ref{fig:jets}a-c) is qualitatively similar to those observed in the open configurations of \citet{Pariat2009,Pariat2010,Pariat2015}. The jet starts at $t_{trig} \approx 570$ and lasts for $t_{jet} \approx 315$ Alfv\'en times. The jet is initiated when the instability onset forcibly reconfigures the sheared configuration, driving the twisted field into the underside of the separatrix dome and generating an extended helical current layer across which magnetic flux reconnects rapidly. The fast reconnection-driven plasma outflows that follow this transition are shown as isosurfaces of velocity magnitude in Fig.\,\ref{fig:jets}; the isosurfaces are colored according to the local value of $v_{z}$, showing the rotational component of the flows as they progress along the loop. Following initiation (Fig.\ \ref{fig:jets}a), as the reconnection continues and the helical current layer forms, the outflow speed increases and takes on a more helical shape (Fig.\ \ref{fig:jets}b). We find that the flow speed peaks at $|\mathbf{v}| \approx 1.0$ in the reconnection exhausts near the null, whereas farther out along the loop typical values drop to $\approx 0.3$. As the current layer rotates around the dome, sequentially reconnecting field lines, it produces a train of torsional Alfv\'{e}n waves that propagate along the loop (Fig.\ \ref{fig:jets}c). Collectively, these waves form a large-scale traveling pulse whose angular velocity vector points along the loop toward the far-loop footpoints, until the leading waves arrive there and begin to reflect back toward the jet source. The reconnection proceeds relatively unhindered by these reflections, as the travel time to the far footpoint and back ($t_{travel}$) in this case is roughly twice the duration of the jet ($t_{jet}$). Thus, we find that the returning flows arrive back at the dome well after the jet ceases.

By comparison, the jet produced in the short loop ($L/N=1.46$) starts much later ($t_{trig} \approx 840$) and is much shorter in duration ($t_{jet} \approx 100$). The later onset can be attributed to a combination of two factors: a significant fraction of the injected shear has already been transferred onto loop field lines via reconnection in the energy storage phase, and more twist is required to initiate the instability that drives the jet. The shorter duration of the jet is due primarily to the shorter travel time along the loop, which causes the reflections from the far loop footpoint to affect the jet source. We estimate the travel time along the loop and back to be $t_{travel} \approx 60$. The interaction of the jet region and the return flows can be seen in Fig.\ \ref{fig:jets}d-f. Following the onset of rapid energy release, fast reconnection-driven plasma flows are launched sequentially toward the apex of the loop, with the leading outflows traveling along the shortest field lines nearest the photosphere (Fig.\ \ref{fig:jets}d). One full travel time after onset, at $t=900$, the previously launched flows along the shortest field lines have reflected off the far loop footpoint and are returning along the loop toward the reconnection region. In the interim, further outflows that curve over the top of this return flow have been launched (Fig.\ \ref{fig:jets}e). Soon after, at $t=920$, the jet outflows become fragmented (Fig.\ \ref{fig:jets}f) as the counterstreaming flows interact. Both fast reconnection and significant magnetic energy release then cease as the reconnection at the dome is choked off by the returning flows.

\subsection{Relaxation Phase}
\label{sec:post-jet}
\begin{figure}[t]
\centering
\includegraphics[width=0.4\textwidth]{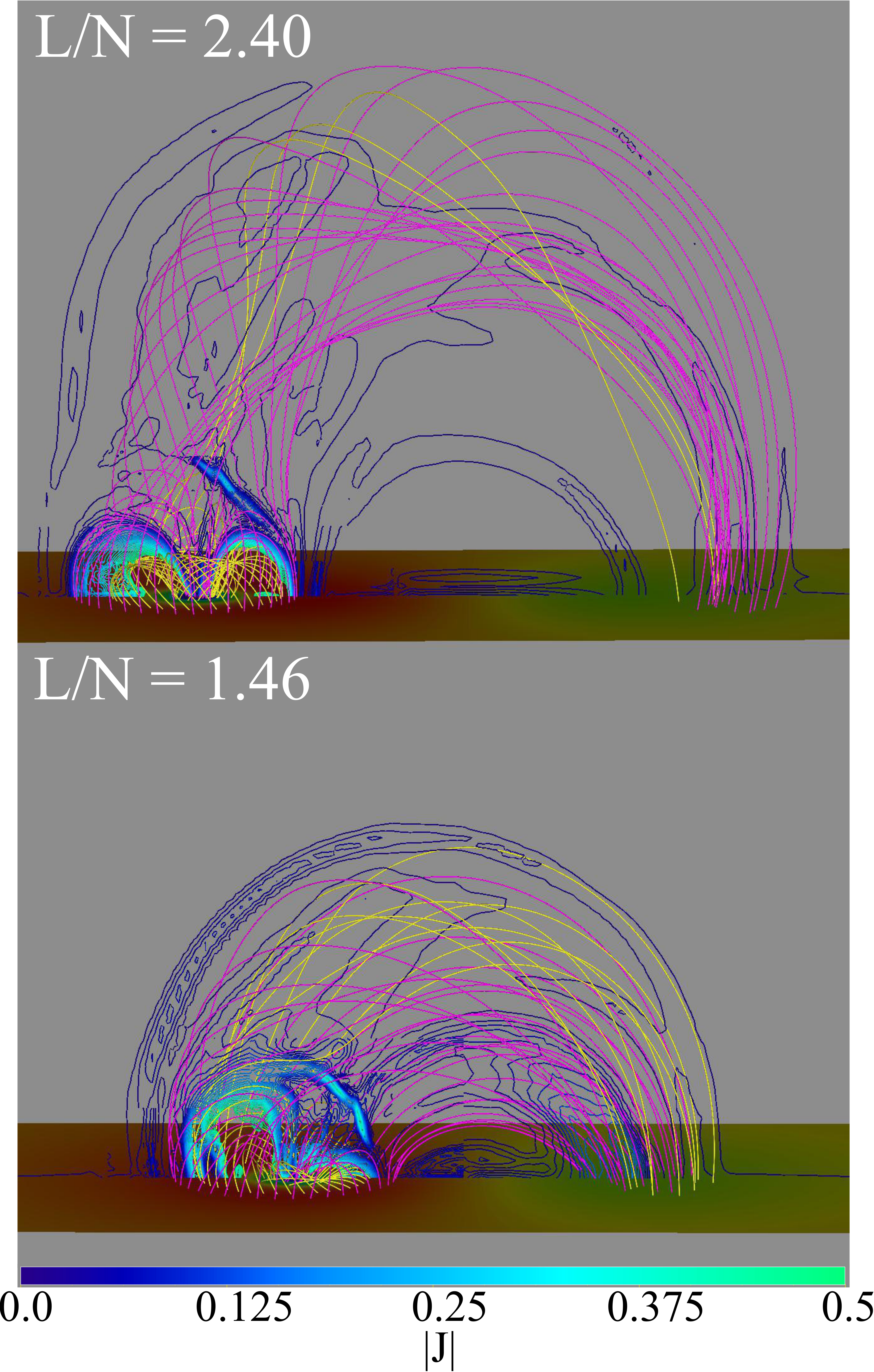}
\caption{The final state ($t=1200$) in the two experiments, with magnetic field lines as in Figure \ref{fig:buildup}. The range and number of $|\mathbf{J}|$ contours has been increased to enhance the lower-amplitude features.}
\label{fig:relax}
\end{figure}

After the main energy release phase has concluded, both loops confining the jets relax toward a new quasi-steady configuration. For $L/N=2.40$, the long loop extends higher into the corona where the field strength is weaker and the loops expand more. As a result, the propagating twist component of magnetic field expands as it reaches the apex of the loop, then narrows again as the disturbance reaches the conjugate footpoint on the photosphere. The fastest flows occur in a curtain-like band around the periphery of the propagating region of twist, where the slingshot effect from the release of magnetic tension is strongest (Fig.\ \ref{fig:jets}c). Counterstreaming flows along the loop are established once the leading wave reflects off of the far loop footpoint. In addition to the torsional waves, a weaker longitudinal oscillation of the loop is generated as the jet propagates along it, so that the loop sways relative to its line-tied footpoints. The torsional waves quickly distribute the injected twist more evenly along the loop as the separatrix dome relaxes (see animation of Fig. \ref{fig:jets}a-c), leaving the twisted loop to oscillate gently as the associated flows gradually damp away. Toward the end of the simulation, weak current layers extend along the expanded coronal loop (Fig.\ \ref{fig:relax}a). These layers reflect mismatches in neighboring flux tube lengths arising from the three-dimensional, inhomogeneous nature of the jet generation. This process has been called reconnection-driven current filamentation \citep{Karpen1996}. A localized current layer also remains at the null, and it continues to slowly release the remnants of twist from beneath the separatrix dome.

For $L/N=1.46$, the short loop with its reduced expansion in height supports similar wave behavior, but increased dissipation from the counterstreaming flows damps the waves more rapidly (see animation of Fig. \ref{fig:jets}d-f). In a manner similar to the long loop, by the end of the simulation a large portion of the injected twist has been redistributed along the loop, and extended current layers permeate the part of the loop affected by the jet. In this case, the currents in these layers are stronger, due to the shorter loop length over which the shear is spread (Fig.\ \ref{fig:relax}b). A current layer also resides at the null in the final state.

\subsection{Reconnection Analysis}
\label{sec:recon1}
To understand more quantitatively how reconnection across the separatrix correlates with the observed jetting behavior, we investigated the connectivity of the magnetic field in the vicinity of the separatrix dome. Using our newly developed field-line integrating routine, we traced field lines from a $500^{2}$ grid of starting positions centered on the parasitic polarity. Each field line either starts and ends beneath the separatrix surface, or starts outside of it and connects to a distant footpoint of the coronal loop. By labeling each starting point accordingly, one can see the footprint of the separatrix surface at a given time. If reconnection occurs across the separatrix over time, this footprint evolves. By tracking the evolution, the associated reconnection rate may also be calculated (see Appendix \ref{ap:recon}). 

Figure \ref{fig:con} shows the footprint of the closed flux beneath the dome in each simulation, at four times: initially (a,e); at the onset of fast reconnection (b,f); at the peak of the reconnection process (c,g); and in the post-jet relaxation phase (d,h). Figure \ref{fig:rr_comp} shows the corresponding calculated reconnection rates. The rapid increase in reconnection rate corresponds closely in each case to the onset of the jet in the volume, whereas the peak in reconnection rate occurs about halfway through the jet (\S \ref{sec:char}).

\begin{figure*}[t]
\centering
\includegraphics[width=0.9\textwidth]{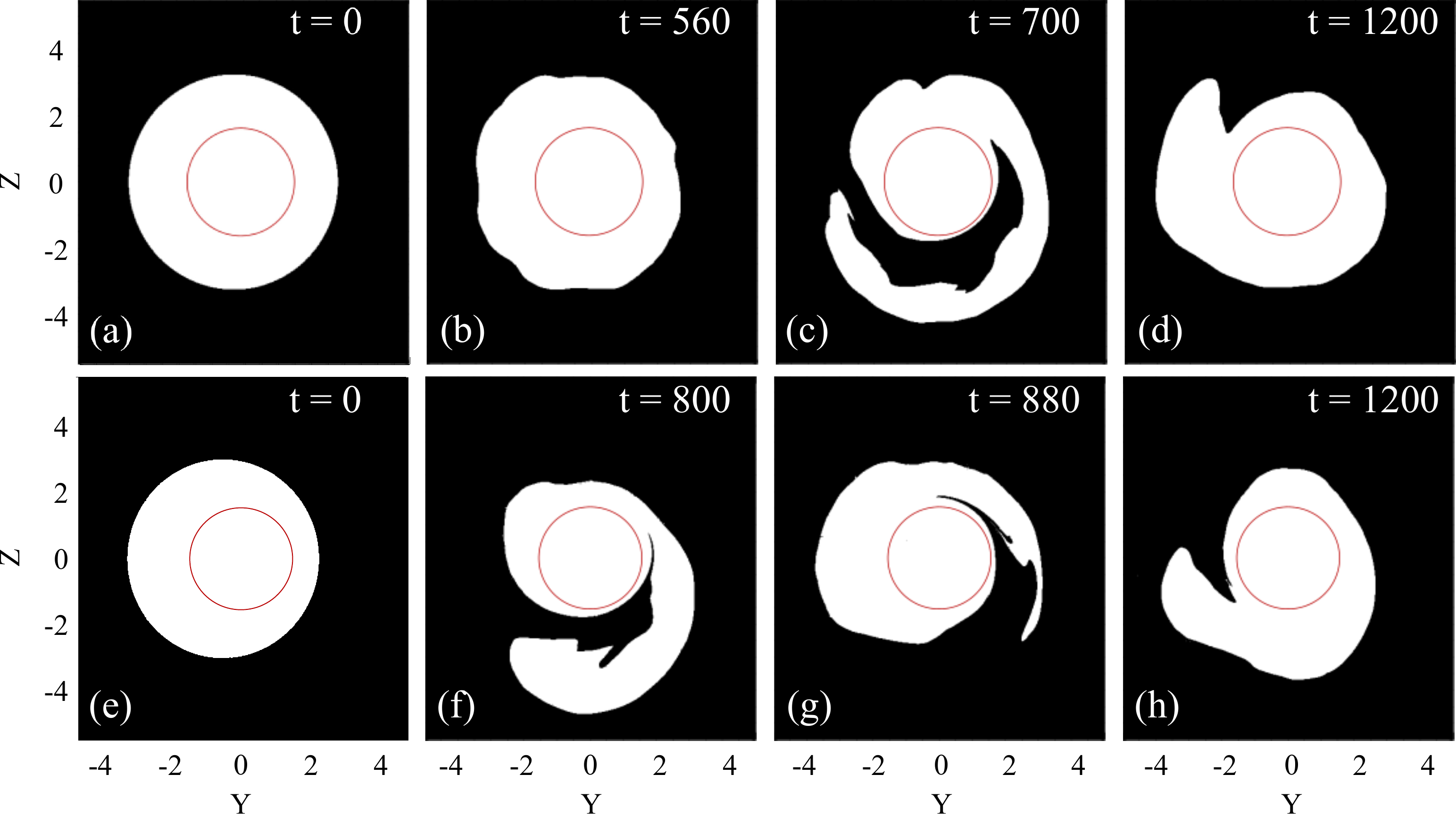}
\caption{Footprint of closed field within the separatrix surface initially (a,e), at the time of fast reconnection onset (b,f), at the time of peak reconnection (c,g), and during the post-jet relaxation (d,f). White regions show magnetic flux that closes beneath the separatrix dome, whereas black regions show flux that connects to the distant coronal-loop footpoints. Red circles show the polarity inversion line in each case. Top panels: $L/N = 2.40$; bottom panels: $L/N = 1.46$. $Y$ and $Z$ are coordinate axes centered on the parasitic polarity. Animations of both configurations are available online.}
\label{fig:con}
\end{figure*}

For $L/N=2.40$, it is clear that the separatrix surface has shifted only slightly during the energy storage phase (Fig.\ \ref{fig:con}a-b), consistent with the low reconnection rate during this period.  At the onset of the kink instability, the reconnection rate increases strongly (Fig.\ \ref{fig:rr_comp}) as the separatrix dome contorts. A channel of locally open field forms, penetrating deep into the previously closed field region as far as the polarity inversion line (Fig.\ \ref{fig:con}c). Once this channel is formed, the footprint of the surface rotates as highly sheared flux near the PIL is reconnected out of the dome and unsheared loop flux is reconnected into it (see the online animation of Fig. \ref{fig:con}). Because the field component normal to the photosphere in these simulations is held constant, the flux beneath the separatrix dome is a fixed quantity. To preserve it, the amount of flux being opened and closed across the separatrix must be equal at any given time \citep[see also][]{Pontin2013}. This type of reconnection is often termed ``interchange reconnection.'' The interchange of a section of highly sheared field with unsheared field reduces the average shear of the closed flux beneath the dome, but does not remove it entirely. The reconnection in the later stages of the jet evolves the field towards a lower energy state that is nearer to potential, and brings the majority of the recently opened (originally closed) flux back beneath the dome. In the final state, the separatrix remains somewhat distorted because the closed field still retains part of the injected twist, whilst the remainder of the twist has been transferred to the enclosing coronal loop. At this stage, reconnection across the separatrix becomes very weak. Overall, the total flux beneath the dome is reconnected roughly twice -- first when the sheared flux is opened, and second when it is closed again -- with the opening and closing occurring principally during the energy-release phase.

By contrast, early reconnection in the more asymmetric configuration for $L/N=1.46$ opens a channel into the sheared field next to the PIL during the energy-storage phase (Fig.\ \ref{fig:con}f). The reconnection rate ramps up at a nearly steady pace during the lead-up to the jet trigger time (Fig.\ \ref{fig:rr_comp}). When the jet is triggered, the separatrix is already highly distorted and roughly all of the flux beneath the dome has been reconnected once (\S \ref{sec:char}). The fast reconnection at the time of the jet (note the spike in Figure \ref{fig:rr_comp}) rapidly closes down the previously opened flux, returning it beneath the separatrix dome. Therefore, in this configuration the ``opening'' of flux occurs much more slowly during the energy-storage phase, whilst only the rapid re-closing of this flux occurs during the energy-release phase. This feature, in addition to the jet reflection along the loop, helps to explain why this jet is so much shorter in duration than that in the long loop.

\begin{figure}[t]
\centering
\includegraphics[width=0.5\textwidth]{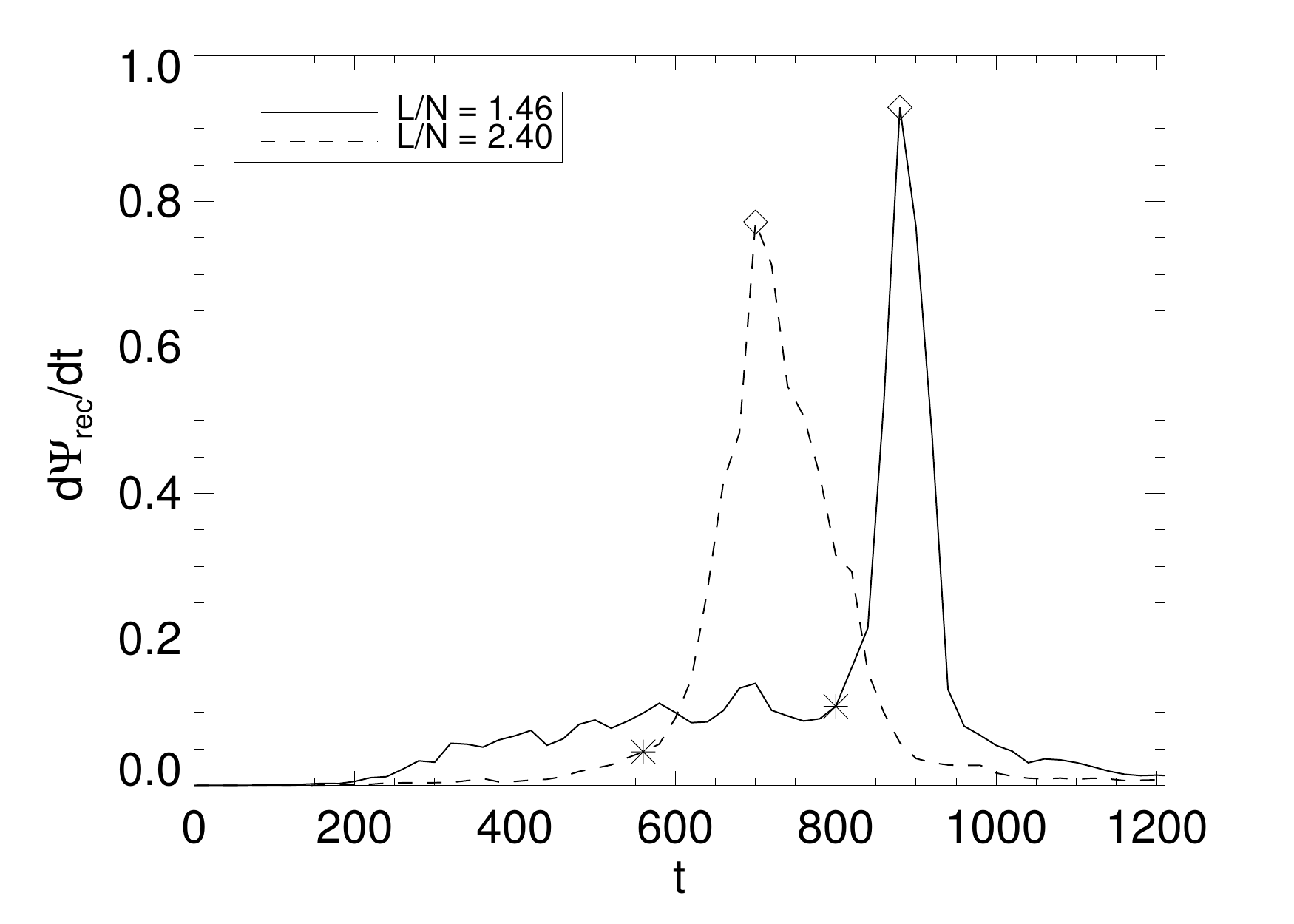}
\caption{Reconnection rates calculated from the flux swept out by the separatrix surface. The times of the connectivity maps shown in Figure \ref{fig:con} are marked by asterisks (reconnection onset; Fig.\ \ref{fig:con}b,f) and diamonds (peak reconnection rate; Fig.\ \ref{fig:con}c,g).}
\label{fig:rr_comp}
\end{figure}

\section{Results II: Parametric Survey}
\label{sec:survey}
The results just described demonstrate that a domed 3D null point topology can produce jets in closed-field configurations. However, the relative size of the jet source region ($N$) with respect to the enclosing coronal loop ($L$) plays an important role in determining the qualitative and quantitative features of the jet. Using the insights gained from the simulations shown above, we now describe the results of the entire parameter study.

\subsection{Initiation}
\label{sec:initiate}
We begin by discussing the conditions for jet onset in the different configurations. All but one of the cases that we studied eventually produced an impulsive jet, although some were very weak. We discuss the exception (with the largest value of $L/N$) at the end of this section. All jets occurred during either the ramp-down phase of the driving or after the driving ceased. We defined the time at which each jet was initiated ($t_{trig}$) as the time when the rate of magnetic-energy liberation suddenly increases (details are discussed below in \S \ref{sec:char}).

\begin{figure}[t]
\centering
\includegraphics[width=0.45\textwidth]{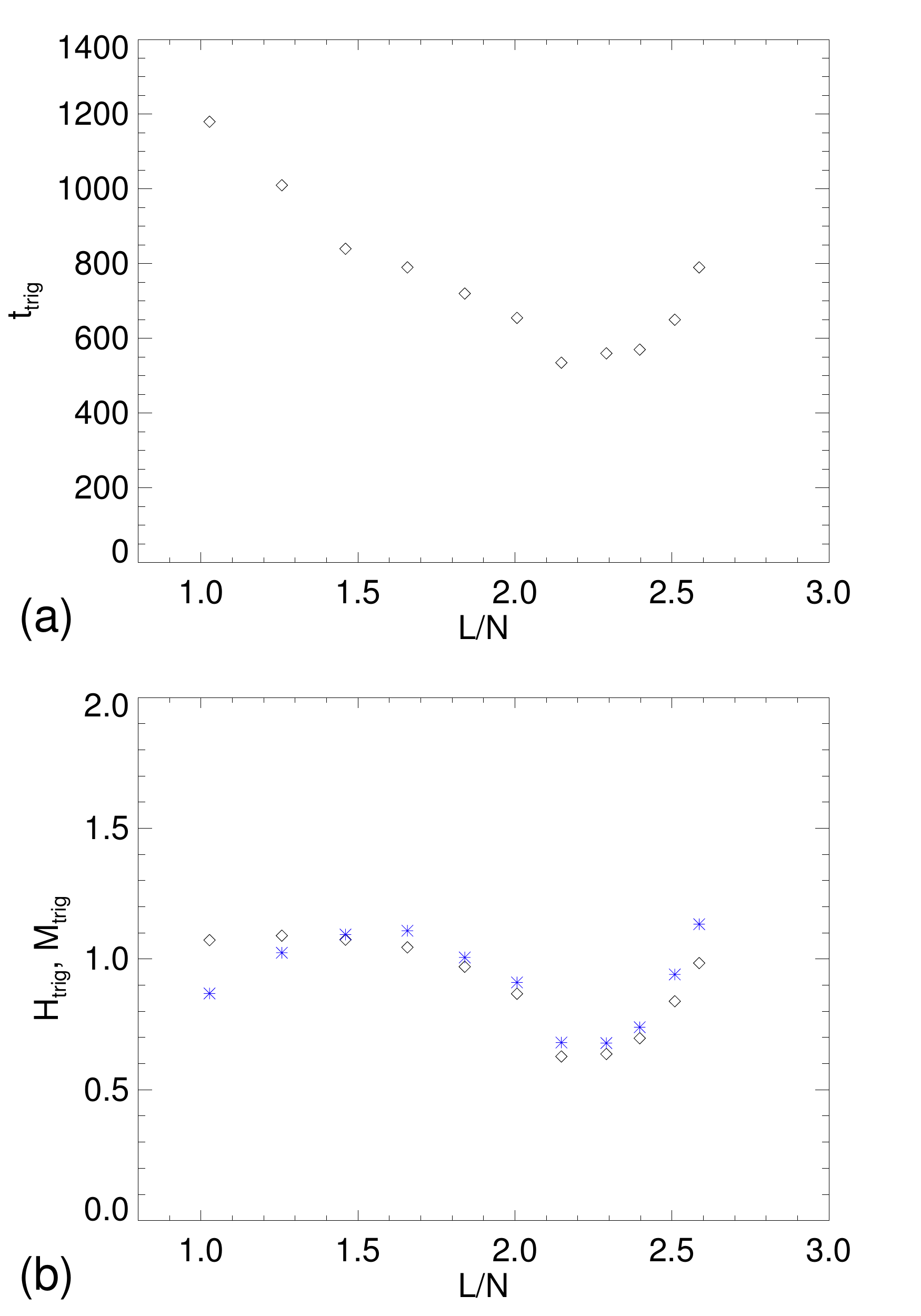}
\caption{For each simulation vs.\ aspect ratio $L/N$: (a) Jet trigger time, $t_{trig}$; (b) At $t=t_{trig}$; injected relative helicity normalized by the square of the flux beneath the dome, $H_{trig}$ (diamonds), and the measured maximum number of turns, $M_{trig}$ (asterisks).}
\label{fig:triggers}
\end{figure}

Figure \ref{fig:triggers}a shows how $t_{trig}$ varies with $L/N$. For the largest values of $L/N$, the null point is nearest the top of the dome and the dome as a whole has the greatest cylindrical symmetry (cf.\ Fig. \ref{fig:buildup}a). This symmetry inhibits the initiation of the kink instability that drives the jet \citep{Pariat2009,Rachmeler2010}. Consequently, as $L/N$ increases the jet is increasingly delayed and $t_{trig}$ increases. For the smallest values of $L/N$, the null point is farthest to the side of the dome and the dome as a whole has the greatest asymmetry (cf.\ Fig. \ref{fig:buildup}d). Current-sheet formation and reconnection occur more readily at the null, reducing the rate of energy buildup beneath the dome (\S \ref{sec:recon1}). This effect delays the jet, so that $t_{trig}$ also increases as $L/N$ decreases. The shortest trigger time occurs at an intermediate value of the aspect ratio, $L/N \approx 2.1$, where the two delaying effects are jointly minimized.

Similar trends are seen in the maximum number of turns, $M$, present in the closed field at $t = t_{trig}$ ($M_{trig} = M(t_{trig})$) and in the normalized relative helicity, $H_{n}$, injected into the volume by this time ($H_{trig}=H_{n}(t_{trig})$), Fig.\ \ref{fig:triggers}b. The former is calculated by counting the number of turns about the center of the parasitic polarity for each of the $500^2$ field lines traced to analyze the magnetic connectivity. The latter quantity is calculated from 
\begin{equation}
H_{n}(t) = \frac{H_{inj}(t)}{\Psi_{dome}^{2}},
\end{equation}
where the injected helicity is  
\begin{align}
H_{inj}(t) &= \int_0^t{\frac{dH_{inj}}{dt'} dt'},\nonumber\\
\frac{dH_{inj}}{dt} &= -2 \int{ \left( {\bf {v}} \cdot {\bf {A}} \right) B_x dS}\\
&= -2 \int{ \int{ \left(v_y A_y + v_z A_z \right) B_x dy dz}}.\nonumber
\end{align}
The last integral, evaluated over the photosphere ($x=0$), is derived from the \citet{Finn1985} gauge-invariant form of the helicity, 
\begin{equation}
H(t) = \int{\left({\bf {A}}+{\bf {A}}_p\right) \cdot \left({\bf {B}}-{\bf {B}}_p\right) dV}.
\end{equation}
Here ${\bf {A}}_p$ and ${\bf {B}}_p$ are the current-free magnetic potential and field, respectively, for the instantaneous $B_x$ distribution at the photosphere. The normalized helicity, $H_{n}$, measures the average twist injected into the flux beneath the separatrix dome. Its value when the jet is initiated, $H_{trig}$, can be seen to match very well the maximum number of turns, $M_{trig}$, measured in the volume numerically at this time. The two deviate for the shortest, most asymmetric systems, where reconnection occurs across the initial separatrix dome prior to jet onset. For $L/N \ge 2.1$, the increasing cylindrical symmetry allows more twist to be stored, increasing the number of turns and the helicity injected into the field before the jet is triggered. For $L/N \le 2.1$, the delayed jet trigger allows more helicity to be injected and more twist to be imparted to the field. However, the increasingly important spine-fan reconnection in this lower range spreads some of the injected helicity along the coronal loop and restricts the strongly twisted field lines to a smaller portion of the flux beneath the dome (\S \ref{sec:recon1}). The helicity eventually plateaus at small $L/N$, whilst the number of turns peaks and then declines again as the reconnection penetrates farther toward the PIL. The turn values that we measured lie within the range $M_{trig} \in [0.7,1.1]$, somewhat below the 1.4 turns required for the initiation of the kink instability in open-field configurations with perfect initial symmetry \citep{Pariat2009,Rachmeler2010} and consistent with the 0.8 turns required in a setup with a tilted background field \citep{Pariat2010}. This supports our interpretation that the impulsive energy-release phase is driven by the onset of a kink-like instability.

Similar results for the variation in jet trigger times for magnetically open configurations have been presented by \citet{Pariat2015}. They investigated the influence of the inclination angle of a straight, uniform, open background field on the initiation and evolution of jets. Their simulations showed the same two effects governing the trigger time of open jets: the kink instability of the twisted field, which occurs later in configurations with more vertical background fields and greater cylindrical symmetry; and spine-fan reconnection across the separatrix, which lowers the rate of energy storage beneath the dome and delays the jet in configurations with more horizontal background fields and greater dome asymmetry. The shortest trigger time occurred for an intermediate inclination angle of the field.

\begin{figure}[t]
\centering
\includegraphics[width=0.45\textwidth]{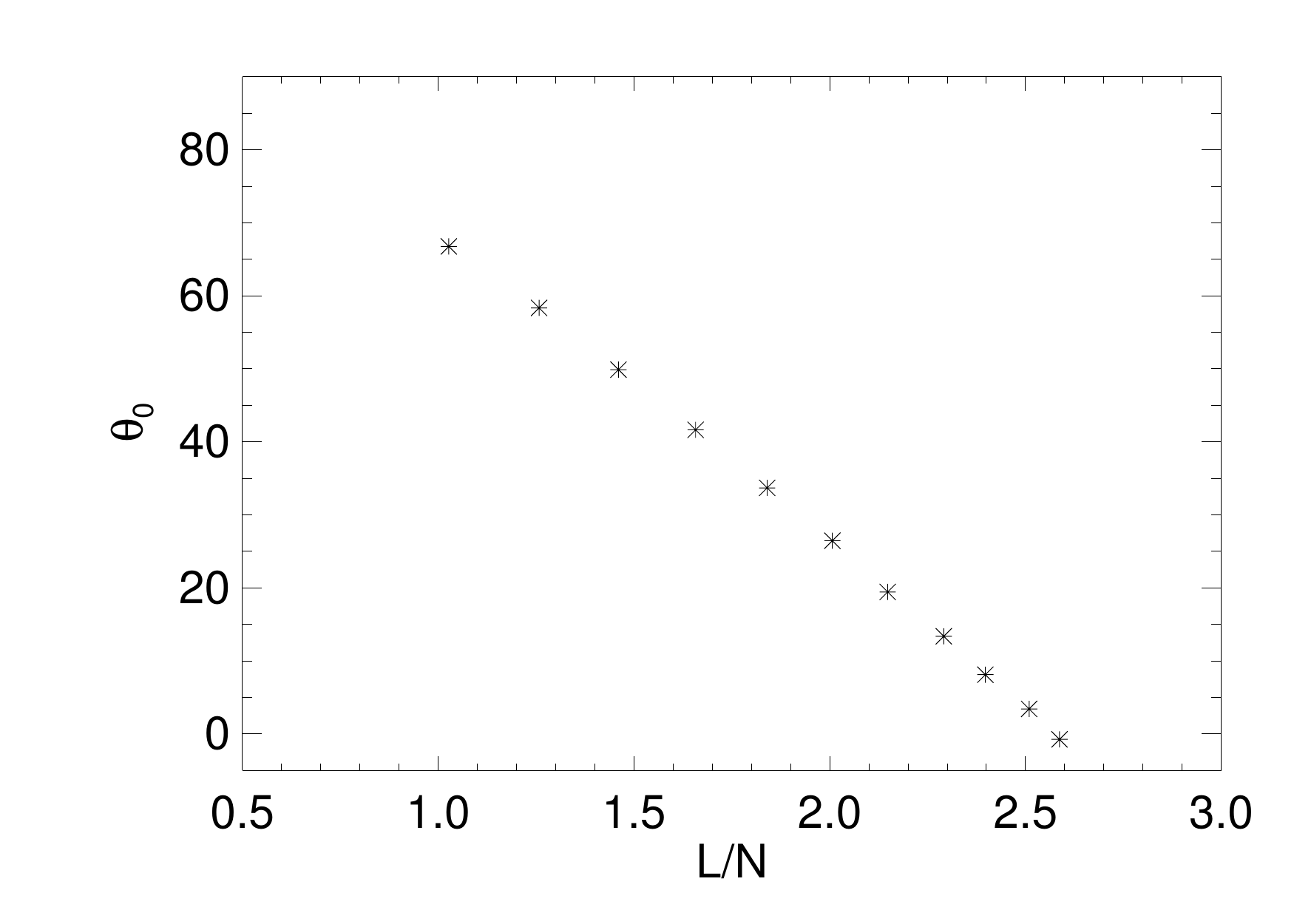}
\caption{For each simulation vs.\ aspect ratio $L/N$: Inclination angle, $\theta_{0}$.}
\label{fig:theta}
\end{figure}

However, there are two important differences between those open-field investigations and our closed-field simulations. First, our background magnetic field falls off with height above the photosphere. Consequently, the separatrix domes in our simulations expand essentially vertically (cf.\ Figs.\ \ref{fig:buildup}c,f), rather than along the direction of the uniform background field as occurs in the open cases. Second, as the dome expands into the closed loop, the tilt angle of the background field at the null increases as the field drapes over the dome. This generates a strengthening non-axisymmetric current layer at the null in all configurations, leading to initially weak reconnection outflows that increase as the dome expands. Therefore, there is no completely axisymmetric configuration as can be achieved in the open-field case, where reconnection is geometrically inhibited during the buildup phase. Given these differences, it is not surprising that the trigger-time-minimizing angle in our simulations differs from the $\approx 8^{\circ}$ tilt angle identified by \citet{Pariat2015}. Although the effective inclination angle in our simulations changes with time, a rough estimate of the initial angle ($\theta_{0}$) can be obtained by using the null position relative to the center of the parasitic polarity at $t=0$. This is a close approximation to the angle of the spine lines near the null, and is equivalent to the angle of the straight background field in open configurations. An approximately inverse linear relationship between $\theta_{0}$ and $L/N$ is found, as shown in Figure \ref{fig:theta}. The aspect ratio that minimizes the jet trigger time for our closed jets, $L/N \approx 2.1$, corresponds to a relatively steep inclination angle, $\theta_{0} \approx 20^{\circ}$.

We conclude this subsection by discussing briefly our longest-loop configuration ($L/N = 2.73$), which remained stable and did not produce an impulsive jet. This case is most cylindrically symmetric, with the null point remaining near the top of the dome throughout the simulation. The number of injected turns ($M_{trig} \approx 1.2$) was less than the critical amount ($M_{trig} \approx 1.4$) needed to set off the kink instability in an open field with zero inclination angle. Draping of field lines, mentioned above, formed a thin current layer surrounding the single null point. The reconnection associated with this current layer was very weak, did not reach the sheared field beneath the dome, and did not destabilize the configuration. We ran this simulation $400$ time units past the end of the driving period, during which time the magnetic field relaxed and reconnection within the current layer tapered off almost entirely. The final state contains a current layer around the null and, evidently, is stable.

\begin{figure*}[t]
\centering
\includegraphics[width=0.9\textwidth]{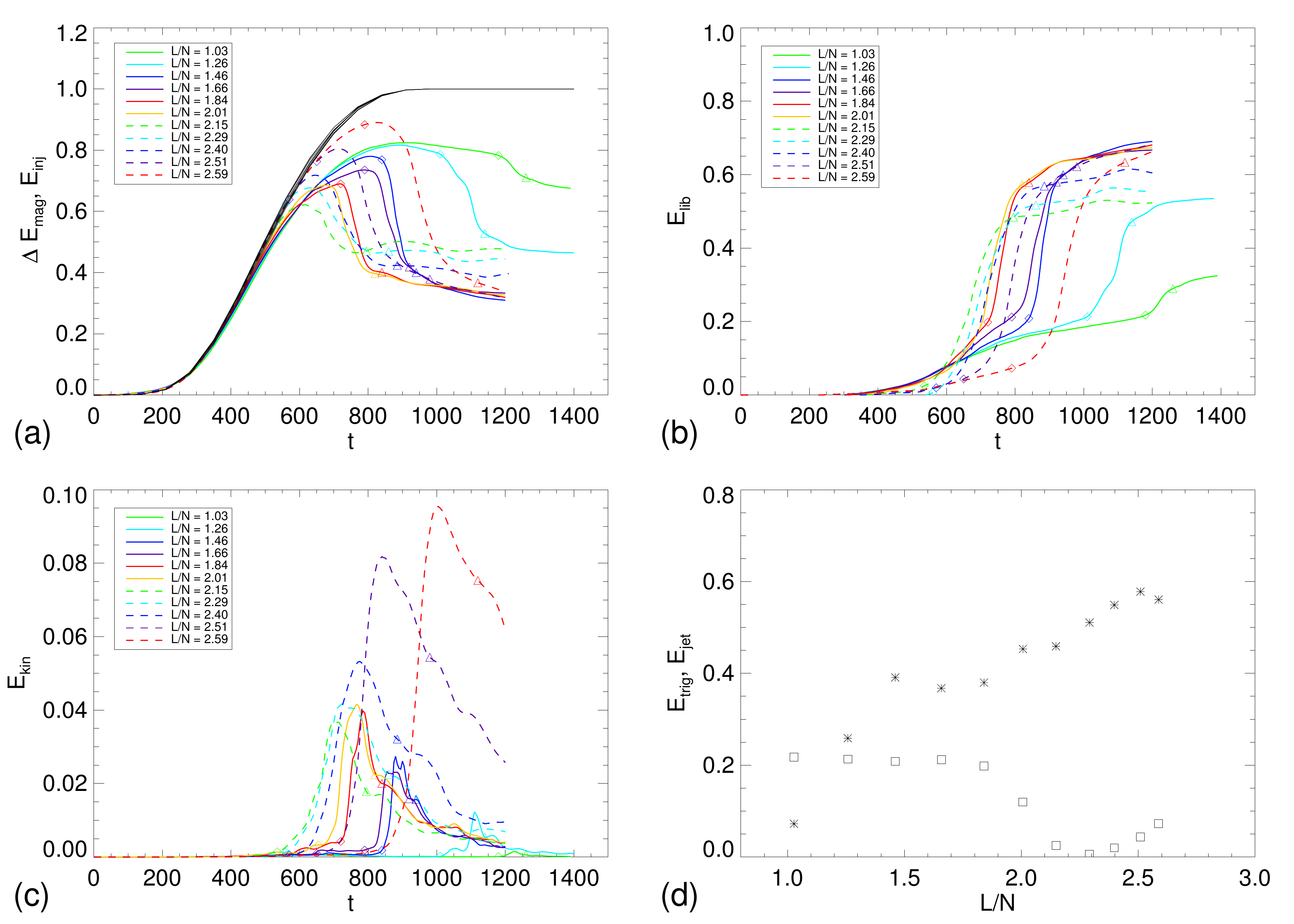}
\caption{Energies normalized to the total injected energy, $E_{inj}^{tot}$, vs.\ time $t$ for each simulation: (a) Injected energy $E_{inj}(t)$ (black lines) and stored magnetic energy $\Delta E_{mag}(t)$ (colored lines); (b) Liberated magnetic energy $E_{lib}(t)$; (c) Kinetic energy $E_{kin}(t)$. In (a)-(c), diamonds and triangles mark the values at $t=t_{trig}$ and $t= t_{trig}+t_{jet}$, respectively. $E_{lib}$ vs.\ aspect ratio $L/N$ for each simulation: (d) Liberated energy up to time $t=t_{trig}$ ($E_{trig}$, boxes) and liberated energy during the energy release phase $0 \le t- t_{trig} \le t_{jet}$ ($E_{jet}$, asterisks).} 
\label{fig:emag}
\end{figure*}

\subsection{Energetics}
\label{sec:char}
We now consider the energetics of our jets. Because each configuration has a slightly different magnetic flux and peak field strength associated with the parasitic polarity, the energy injected by the boundary driving is different in each case. The total energy injected into the magnetic field by the footpoint motions up to time $t$ is calculated by integrating the Poynting flux across the bottom boundary:
\begin{align}
E_{inj}(t)&= \int_0^t{\,dt' \int{c \left( \mathbf{E}\times\mathbf{B} \right) \cdot \mathbf{n} \,dS}} \nonumber\\
&= -\int_0^t{\,dt' \int{\int{\left( v_{y}B_{y}+v_{z}B_{z} \right) B_{x}\,dy}\,dz}}.\nonumber \\
\end{align}
To compare the runs equitably, we normalize the various energies with respect to the total energy injected into the volume by the footpoint motions in each case, $E_{inj}^{tot} = E_{inj}(t \rightarrow \infty)$. Table \ref{table:param} lists the resulting values of $E_{inj}^{tot}$ for each configuration.

Early in each simulation, essentially all of the injected energy is stored in the coronal magnetic field, as very little is converted to kinetic or thermal energy of the plasma. Because the component of the magnetic field normal to the photosphere is held fixed throughout each run, the current-free field ${\bf{B}}_p$ and its energy $E_p$ also are independent of time. Thus, the instantaneous free magnetic energy is given simply by 
\begin{equation}
\Delta E_{mag}(t) = E_{mag}(t) - E_{mag}(0).
\end{equation}
Figure \ref{fig:emag}a shows the evolution of the normalized $\Delta E_{mag}(t)$ in each of our jet-producing simulations. Also shown is the profile of $E_{inj}(t)$, which after normalization is nearly identical for all cases. The onset of the jet in each configuration is signaled by a rapid decrease in $\Delta E_{mag}$. As previously mentioned, jet onset occurs over a broad range of times with respect to the phase of the boundary driving: some occur during the ramp-down phase ($500 < t < 1000$), others after the driving ceases ($t \ge 1000$). To characterize the time and duration of each jet, we focused on the cumulative energy liberated from the magnetic field,
\begin{equation}
E_{lib}(t) = E_{inj}(t) - \Delta E_{mag}(t).
\end{equation}
This quantity, shown in Figure \ref{fig:emag}b after being normalized to $E_{inj}^{tot}$, measures the total energy released by the magnetic field whilst taking into account the different absolute energy injection rates at the different times of jet onset. We define the trigger time, $t_{trig}$, as the time when a noticeable increase in $E_{lib}$ occurs (corresponding to fast energy release as the jet begins) and the jet duration, $t_{jet}$, as the time elapsed thereafter until the sharp increase in $E_{lib}$ subsides. Table \ref{table:param} lists $t_{trig}$ and $t_{jet}$ in each simulation. Diamonds and triangles in Figure \ref{fig:emag}b mark the normalized $E_{lib}$ values at times $t_{trig}$ and $t_{trig}+t_{jet}$, respectively. These times are well-correlated with the rapid increase and subsequent decline of the volumetric kinetic energy, $E_{kin}(t)$, following the launch, travel, and deceleration of the jet in each case (Fig.\ \ref{fig:emag}c).

The transition in behavior around $L/N=2.1$, from jets preceded by significant reconnection in the energy buildup phase to jets with relatively little, is evident in the magnetic energy curves. In Figure \ref{fig:emag}a, the injected energy closely matches the free magnetic energy prior to the jet ($t<t_{trig}$) for configurations with $L/N > 2.1$ (dashed lines), due to the weak reconnection occurring in this phase. Correspondingly, the liberated energy $E_{lib}$ is small until the onset of the impulsive phase of these jets, when large increases in $E_{lib}$ are evident (Fig.\ \ref{fig:emag}b). In contrast, for $L/N < 2.1$, the injected and free energy curves deviate early in the evolution (Fig.\ \ref{fig:emag}a, colored solid lines), with significant fractions of the injected energy liberated prior to jet onset in the most asymmetric (smallest $L/N$) configurations (Fig.\ \ref{fig:emag}b). Despite the marked differences in the timing of energy liberation among the various cases, a rather consistent total of $50\%$ to $60\%$ of the injected energy is liberated during the combined energy-storage and -release phases (Fig.\ \ref{fig:emag}b).

Figure \ref{fig:emag}d shows that the energy liberated by time $t = t_{trig}$,
\begin{equation}
E_{trig} = E_{lib}(t_{trig}),
\end{equation}
normalized to $E_{inj}^{tot}$, is as large as $20\%$ for the cases with small ratios of $L/N$ and drops to less than $5\%$ for configurations with $L/N > 2.1$. The slight increase at the largest values of $L/N$ is due to late-time, weak reconnection associated with draping of field lines over the strongly expanded dome. The energy liberated during the impulsive energy-release phase itself,
\begin{equation}
E_{jet} = E_{lib}(t_{trig}+t_{jet}) - E_{lib}(t_{trig}), 
\end{equation}
normalized to $E_{inj}^{tot}$, increases from about $40\%$ for moderate ratios of $L/N$ to $60\%$ for the largest. There is a sharp fall-off in this quantity at small aspect ratios ($L/N<1.3$), due to the significant reconnection in the energy storage phase and the jet reflection along the short loops for these highly asymmetric cases. The impaired energy release produces weak jets with small kinetic energies (Fig.\ \ref{fig:emag}c).

\begin{figure*}[t]
\centering
\includegraphics[width=0.9\textwidth]{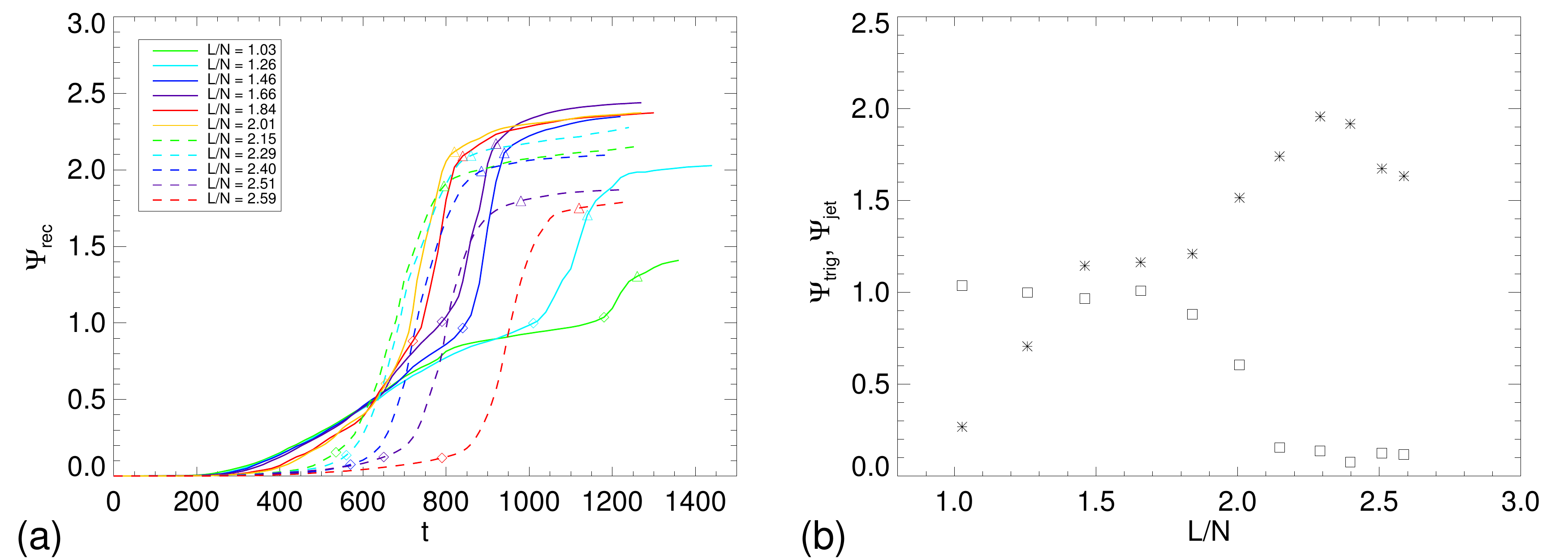}
\caption{Cumulative reconnected fluxes, $\Psi_{rec}$, normalized to the flux beneath the separatrix dome, $\Psi_{dome}$, for each simulation: (a) Evolution vs.\ time $t$, with values marked when $t=t_{trig}$ (diamonds) and $t= t_{trig}+t_{jet}$ (triangles); (b) Variation vs.\ aspect ratio $L/N$ at time $t=t_{trig}$ ($\Psi_{trig}$, boxes) and during the energy-release phase $0 \le t-t_{trig} \le t_{jet}$ ($\Psi_{jet}$, asterisks).}
\label{fig:recon}
\end{figure*}

\begin{figure*}[t]
\centering
\includegraphics[width=0.9\textwidth]{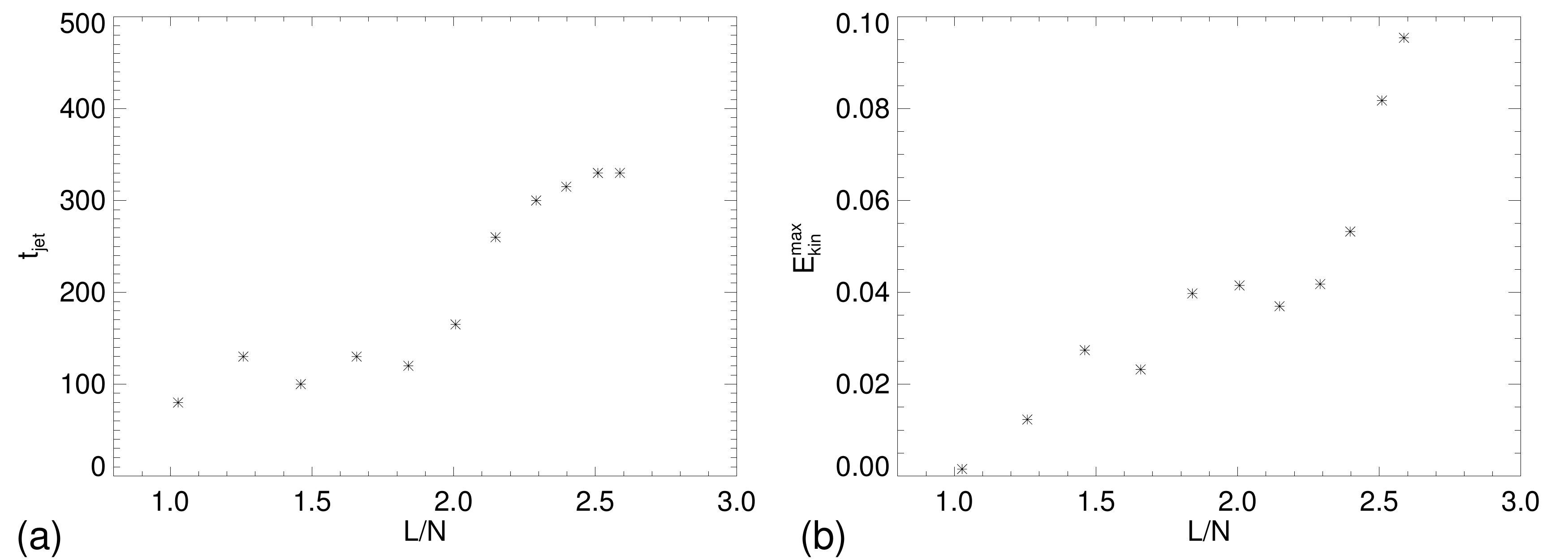}
\caption{For each simulation vs.\ aspect ratio $L/N$: (a) Jet duration, $t_{jet}$; (b) Peak kinetic energy, $E_{kin}^{max}$, normalized to $E_{inj}^{tot}$.}
\label{fig:flows}
\end{figure*}

A deeper understanding of the energetics of our jets results from considering the cumulative reconnected flux, normalized to the flux within the separatrix dome, shown in Figure \ref{fig:recon}a. The resemblance between the curves of free-energy liberation (Fig.\ \ref{fig:emag}b) and of interchange-reconnected flux across the separatrix (Fig.\ \ref{fig:recon}a) is striking. Close similarities can also be seen between the magnetic flux reconnected prior to each jet,
\begin{equation}
\Psi_{trig} = \Psi_{rec}(t_{trig}),
\end{equation} 
and during the energy-release phase,
\begin{equation}
\Psi_{jet} = \Psi_{rec}(t_{trig}+t_{jet}) - \Psi_{rec}(t_{trig}),
\end{equation} 
normalized to $\Psi_{dome}$, shown in Figure \ref{fig:recon}b, with the corresponding liberated energies over these time intervals (Fig.\ \ref{fig:emag}d). Both show the direct link between the rates of interchange reconnection and energy release prior to and during the jet.

By the end of the energy-release phase, all but the most compact, asymmetric configurations reconnect the flux beneath the dome at least twice ($\Psi_{rec} \ge 2\Psi_{dome}$): once when the sheared field near the PIL reconnects to distribute twist along the coronal loop, and again when it reconnects back down beneath the dome. However, as discussed in \S \ref{sec:recon1}, for small $L/N$ the opening occurs during the energy-storage phase, whilst for large $L/N$ the opening and closing are delayed until the energy-release phase. As was found for other characteristic jet quantities, the transition between these two behaviors of the reconnected flux occurs at $L/N \approx 2.1$.

These trends in the energy liberated and the flux reconnected also are reflected in the durations of the jets and in the associated peak kinetic energies (Fig.\ \ref{fig:flows}a,b). Those configurations with large values of $L/N$, which reconnected the most flux and liberated the most free energy during the energy-release phase, produced the longest-lived jets with the greatest kinetic energies (up to $\approx 10\%$ of the total injected energy), whilst those with the smallest ratios of $L/N$ produced the shortest-lived, least-energetic jets. The duration almost triples in length, from $t_{jet} \approx 120$ to $\approx 340$, with a steep transition around $L/N=2.1$ (Fig.\ \ref{fig:flows}a), correlated with the reduction in normalized flux reconnected prior to jet onset (Fig.\ \ref{fig:recon}b; boxes). The peak kinetic energies show a more complex dependence upon the aspect ratio $L/N$, but generally increase for larger ratios. Jets in shorter loops (small $L/N$)  have significantly shorter travel times along the loop ($t_{travel}$) than the duration of the jet itself ($t_{jet}$), sometimes yielding double-peaked kinetic energy curves as the counterstreaming flows interfere with each other. For higher values of $L/N$ the travel time increases, but this is mediated in part by the increase in jet duration. By inspection, we find that jets with $L/N\ge 1.7$ are relatively unaffected by reflections from the far loop footpoints, although they do exhibit counterstreaming flows in the loop.

\section{Correspondence with Observed Jets}
\label{sec:obs0}

\subsection{Qualitative Features}
\label{sec:obs1}

In all of the configurations that we studied, the jet produced was confined by a coronal loop whose footprint envelopes the jet source region. This is a generic feature of all closed-field jets, most evident when the coronal loop has significant curvature \citep[e.g.][]{Shibata1992,Shimojo1998,Torok2009,Yang2012,Guo2013,Lee2013,Schmieder2013,Zheng2013,Cheung2015}. The curtain-like shape of our jets was shown to expand as it propagates towards the loop top (driven by the expansion of the ambient field as the background field strength declines with height) and contract again toward the far loop footpoint. This effect is expected to be most observable in large jets that have a relatively wide source region, so that the affected loop strands expand noticeably in the corona and the jet flows are energetic enough to defy gravity and reach high up along the loop. The animation of the active region jet in Fig.\ \ref{fig:sdoaia} shows such a large jet where this expansion and contraction are clearly visible.

Our model predicts differing behaviors for the waves launched during the impulsive jet phase depending upon the duration of the jet ($t_{jet}$) compared with the travel time along the coronal loop ($t_{travel}$). When $L/N$ is very large, $t_{travel} \gg t_{jet}$ and the jet-launched wave motions are expected to travel freely along the coronal loops. \citet{Torok2009} observed and modeled an example of such freely propagating torsional wave motions along longer coronal loops. When the two time scales are comparable, our model predicts that the jet generation is relatively unhindered by reflections from the far loop footpoint, but counterstreaming flows will be present in the loop. A possible manifestation of this was described by \cite{Qiu1999}, who observed counterstreaming flows in H$\alpha$ loops in an active region at the same time that flaring and jetting were observed in a mixed-polarity region at one end of the loop system. When $L/N$ is small and $t_{travel} < t_{jet}$, our model predicts that the reflected jet-launched waves will interfere with the further development of the jet. The interaction of the returning flows and the jet outflows could conceivably generate turbulence in the loop, leading to extended emission along the connecting loops as well as within the anemone region. \citet{Shimojo1998} described soft X-ray emission spread along short coronal loops where the estimated travel time was shorter than the jet lifetime, lending some credence to this idea.

As in open-field regions, closed-field jets sometimes are generated repeatedly from the same source region \cite[e.g.][]{Cheung2015}. The free energy that drives the jets in our model arose from boundary motions, applied on the photosphere, which were stopped once a single jet was produced. In open fields, \citet{Pariat2010} showed that homologous jets are generated by maintaining the driving. We tested this on one of our simulations by maintaining the photospheric driving, and found that it too produced homologous jets along the same coronal loop. Thus, our model can explain the observed homology of some closed-field jet regions.

\subsection{Quantitative Measures}
\label{sec:obs2}
Quantitatively, the properties of our jets can be compared to those observed by applying scale factors $\rho_s$, $B_s$, and $L_s$ to the dimensionless simulated quantities to obtain typical coronal values of mass density, magnetic field strength, and length, respectively. The associated scale factors for pressure ($P_s$), velocity ($V_s$), time ($t_s$), energy ($E_s$), and magnetic reconnection rate (flux per unit time; $\dot{\Psi}_s$) are 
\begin{align}
P_s &= B_s^2, \nonumber\\
V_s &= \frac{B_s}{\sqrt{\rho_s}}, \nonumber\\
t_s &= \frac{L_s}{V_s}, \\
E_s &= B_s^2 L_s^3, \nonumber\\
\dot{\Psi}_s &= B_s V_s L_s. \nonumber
\end{align}
For simplicity, in this discussion we will work solely with orders of magnitude in the scale factors.

The most elementary example is jets occurring in areas of quiet Sun ($qs$).  There, we can assume $B_s = 1$, hence the strengths of the coronal-loop ($B_{cl}$) and parasitic-polarity ($B_{pp}$) fields are
\begin{equation}
B_{cl}^{qs} \approx 4~{\rm G},~~~B_{pp}^{qs} \approx 21~{\rm G}.
\end{equation}
The pressure scale factor $P_s = 1$, so the thermal pressure ($P_{th}$) is 
\begin{equation}
P_{th}^{qs} \approx 1 \times 10^{-2}~{\rm dyn~cm}^{-2}.
\end{equation}
At a temperature of $1 \times 10^6$ K, the corresponding mass density is
\begin{equation}
\rho^{qs} \approx 1 \times 10^{-16}~{\rm g~cm}^{-3}.
\end{equation}
Hence, in quiet Sun we set $\rho_s = 1 \times 10^{-16}$, which together with $B_s = 1$ gives $V_s = 1 \times 10^8$. Our peak dimensionless jet velocity is $\approx 1.0$, so we find
\begin{equation}
V_{jet}^{qs} \approx 1 \times 10^{8}~{\rm cm~s}^{-1}.
\end{equation}
This is consistent with the maximum apparent jet flow speeds reported by \citet{Shimojo1996} and \citet{Savcheva2007}. Typical velocities within our main curtain-like spray were somewhat lower, with dimensionless values $\approx 0.3$, scaling to $3\times 10^{7}$ cm s$^{-1}$. This falls within the ranges of velocities reported by \citet{Shimojo1996} and \citet{Savcheva2007}, and is consistent with the jet speeds reported by \citet{Shibata1992}. Finally, a length scale factor $L_s = 1 \times 10^9$ applied to our average dimensionless loop length gives a coronal value 
\begin{equation}
L^{qs} \approx 12L_s \approx 1.2 \times 10^{10}~{\rm cm}.
\end{equation}
For this long coronal loop, the remaining scale factors are $t_s = 1 \times 10^1$, $E_s = 1 \times 10^{27}$, and $\dot{\Psi}_s = 1 \times 10^{17}$. Using values from Table \ref{table:param} for the typical jet duration, peak kinetic energy, and peak reconnection rate we obtain, respectively,
\begin{align}
t_{jet}^{qs} &\approx 200 t_s \nonumber\\
&\approx 2 \times 10^3~{\rm s}, \\
E_{kin}^{qs} &\approx .04 \times 125 E_s \nonumber\\
&\approx 5 \times 10^{27}~{\rm erg}, \\
\dot{\Psi}_{rec}^{qs} &\approx 1.0 \dot{\Psi}_s \nonumber\\
&\approx 1.0 \times 10^{17}~{\rm Mx~s}^{-1}.
\end{align}
The duration of our quiet-Sun jet is roughly the mean of the large range of jet lifetimes reported by \citet{Shimojo1996} ($\approx 2 \times 10^2$--$2 \times 10^4$ s) and is at the high end of the range reported by \citet{Savcheva2007} ($\approx 2 \times 10^2$--$2 \times 10^3$ s). Its kinetic energy is near the upper end of the range reported by \citet{Shibata1992} ($\approx 1 \times 10^{28}$ erg). The reconnection rate varies from about $1 \times 10^{16}$ Mx s$^{-1}$ during the early, slow-reconnection (energy-storage) phase to $1 \times 10^{17}$ Mx s$^{-1}$ during the impulsive, fast-reconnection (energy-release) phase of the jets.

For jets occurring in active regions ($ar$), we set $B_s = 10$. The strengths of the coronal-loop ($B_{cl}$) and parasitic-polarity ($B_{pp}$) fields then are
\begin{equation}
B_{cl}^{ar} \approx 40~{\rm G},~~~B_{pp}^{ar} \approx 210~{\rm G}.
\end{equation}
The pressure scale factor $P_s = 100$, so now the thermal pressure is 
\begin{equation}
P_{th}^{ar} \approx 1 \times 10^0~{\rm dyn~cm}^{-2}.
\end{equation}
Here, the corresponding mass density is
\begin{equation}
\rho^{ar} \approx 1 \times 10^{-14}~{\rm g~cm}^{-3}.
\end{equation}
We therefore set $\rho_s = 1 \times 10^{-14}$, which together with $B_s = 10$ again gives $V_s = 1 \times 10^8$. Thus, as in quiet Sun we find
\begin{equation}
V_{jet}^{ar} \approx 1 \times 10^{8}~{\rm cm~s}^{-1}.
\end{equation}
Here, we assume a length scale factor $L_s = 1 \times 10^8$ to model jets in compact active-region loops, 
\begin{equation}
L^{ar} \approx 12L_s \approx 1.2 \times 10^{9}~{\rm cm}.
\end{equation}
The rest of the scale factors are $t_s = 1$, $E_s = 1 \times 10^{26}$, and (as before) $\dot{\Psi}_s = 1 \times 10^{17}$. For the typical jet duration, peak kinetic energy, and peak reconnection rate we obtain, respectively, 
\begin{align}
t_{jet}^{qs} &\approx 200 t_s \nonumber\\
&\approx 2 \times 10^2~{\rm s}, \\
E_{kin}^{qs} &\approx .04 \times 125 E_s \nonumber\\
&\approx 5 \times 10^{26}~{\rm erg}, \\
\dot{\Psi}_{rec}^{qs} &\approx 1.0 \dot{\Psi}_s \nonumber\\
&\approx 1.0 \times 10^{17}~{\rm Mx~s}^{-1}.
\end{align}
The duration of our active-region jet is at the low end of the range of jet lifetimes reported by \citet{Shimojo1996} and \citet{Savcheva2007}, and somewhat shorter than the observed jet shown in Figure \ref{fig:sdoaia}. Its kinetic energy is near the mean of the range reported by \citet{Shibata1992} ($\approx 1 \times 10^{25}$ -- $1 \times 10^{28}$ erg). The reconnection rate varies over the same range as in quiet Sun, $1 \times 10^{16}$ Mx s$^{-1}$ to $1 \times 10^{17}$ Mx s$^{-1}$.

\section{Discussion}
\label{sec:discuss}
We have investigated the initiation and evolution of solar jets in closed coronal loops via three-dimensional numerical simulations of the embedded-bipole model \citep{Antiochos1996}, which has been investigated extensively to explain coronal-hole jets \citep{Pariat2009,Pariat2010,Pariat2015}. As in those prior studies, our impulsive jets consist of twisted, curtain-like outflows that exhibit strong helical motions and filamentary structure, resembling observations \citep[e.g.][]{Torok2009,Yang2012,Guo2013,Lee2013,Schmieder2013,Zheng2013,Cheung2015}. Due to the variation in loop cross-section with distance along the loop, our jets expand and then contract laterally as they propagate along the loop toward its far footpoint, as observed. By scaling our dimensionless results using typical input parameters characterizing coronal jet sources \citep{Shibata1992,Shimojo1996}, we obtain values for the jet speeds, durations, and energies that are consistent with those observed (\S \ref{sec:obs2}).

Our results further show that the evolution is highly sensitive to the relative sizes of the closed dome ($N$) of the jet source region and the coronal loop ($L$) within which the source is embedded. We found that configurations with large $L/N$ ratios store the greatest amount of magnetic free energy and produce the longest-duration, most energetic jets; those with small $L/N$ ratios release a significant fraction of the injected free energy prior to onset of their shorter-duration, less energetic jets. The transition between these behaviors occurs at $L/N \approx 2.1$, where the number of turns of induced twist required to initiate the impulsive jet is minimized (Fig.\ \ref{fig:triggers}b). Smaller configurations liberate up to 20\% of the stored free energy prior to jet onset; larger ones liberate essentially none (Fig.\ \ref{fig:emag}d). These energies reflect amounts of cumulative reconnected flux prior to onset of as much as 100\% and as little as 10\%, respectively, of the total flux enclosed beneath the dome (Fig.\ \ref{fig:recon}b). During the fast reconnection that drives the impulsive jet, in contrast, the liberated energy ranges from less than 10\% for small $L/N$ to as much as 60\% for large $L/N$ (Fig.\ \ref{fig:emag}d), and the reconnected flux ranges from as little as 30\% to as much as 200\% (Fig.\ \ref{fig:recon}b). The corresponding jet durations range from fewer than 100 to more than 300 Alfv\'en times (Fig.\ \ref{fig:flows}a). Finally, the kinetic energies in the jet flow, normalized to the total injected energy, range smoothly from essentially zero at $L/N \approx 1$ to about 5\% at $L/N \approx 2.4$, and then rise at a more rapid pace toward higher $L/N$ (Fig.\ \ref{fig:flows}b). In configurations with $L/N > 2.4$, the reconnection driving the impulsive jet ceases before the jet flow reaches the far footpoint of the loop and reflects back into the jet source region. In configurations with $L/N < 1.7$, on the other hand, the counterstreaming forward and backward (reflected) jet flows interfere strongly with each other (Fig.\ \ref{fig:jets}), and even appear to choke off the reconnection outflow in our most compact configurations.

Following the cessation of the impulsive reconnection and subsidence of the principal jet outflows, our system relaxes toward a new quasi-steady state with filamentary current structures threading the coronal loop. These structures arise from a local mismatch of neighboring flux-tube lengths, driven by the spatially and temporally intermittent transfer of twist to the loop during the three-dimensional evolution: reconnection-driven current filamentation \citep{Karpen1996}. At higher grid resolution, these currents should become even more filamentary in structure and greater in strength, and could produce quasi-steady heating of the loop in the aftermath of the jet. In addition, it is known that 3D null-point current layers are explosively unstable to resistive tearing at high Lundquist numbers \citep{Wyper2014a,Wyper2014b}. Attaining the requisite Lundquist-number threshold demands better resolution than the simulations undertaken in this work. Sufficiently high-resolution simulations also should reveal more fine structure in the jet itself, plausibly including the formation and ejection of small-scale plasma concentrations \citep[``blobs'';][]{Zhang2014} in the jet outflow. These intriguing possibilities are being evaluated in our ongoing study of coronal loop jets.

\acknowledgments
This work was supported by P.F.W.'s appointment to the NASA Postdoctoral Program, administered by Oak Ridge Associated Universities through a contract with NASA, and by C.R.D.'s participation with a NASA Living With a Star Focused Science Team on solar jets. The computer resources used to perform the numerical simulations were provided to C.R.D. by NASA's High-End Computing program at the NASA Center for Climate Simulation. Fig. \ref{fig:sdoaia} and its animation were created using the ESA and NASA funded Helioviewer Project. We are grateful to our colleagues Spiro Antiochos, Judy Karpen, Etienne Pariat, and Kevin Dalmasse for numerous helpful discussions on the topic of solar jets, to David Pontin for ongoing discussions and insight regarding three-dimensional reconnection, and to our anonymous referee for suggesting clarifying changes to the original manuscript.

\appendix
\section{Reconnected Flux and Reconnection Rate}
\label{ap:recon}
To calculate the reconnection rate in our simulations we take advantage of the fact that there is a true separatrix surface across which the rate of flux transfer may be measured. In each of the simulations the boundary driving is localized within the circular polarity inversion line so that near the separatrix the field line footpoints are line tied and fixed in position. Any change in the position of the separatrix is therefore due to reconnection occurring within the volume. Since the driven ends of the closed field lines always remain beneath the dome we can ignore the fact that these field lines are moving relative to the position of the starting grid from which the field lines are traced. 

We calculate the instantaneous reconnection rate in the following manner. For each field line in our tracing grid we assigned a magnetic flux element
\begin{equation}
\Delta \Psi_{i,j} = (B_{n})_{i,j} \,\Delta y\, \Delta dz,
\end{equation}
where $\Delta y$ and $\Delta z$ are the separation of the starting positions in the grid and $(B_{n})_{i,j}$ is the magnetic field normal to the photosphere at this position. At each time we compare the connectivity of each field line with the previous time. The flux elements of all the field lines which have been ``opened'' and ``closed'' in this time interval are summed to give the total opened and closed flux respectively
\begin{equation}
\Delta \Psi_{opened} = \sum_{opened~i,j} \Delta \Psi_{i,j}, \quad \Delta \Psi_{closed} = \sum_{closed~i,j} \Delta \Psi_{i,j}.
\end{equation}
To a high degree of accuracy in each of our simulations, the opened flux matched the closed flux. The reconnection rate is taken to be the simple average 
\begin{equation}
\dot{\Psi}(t_k) = \frac{\Delta \Psi_{opened}+\Delta \Psi_{closed}}{2 \Delta t_k}.
\end{equation}
Here $t_k$ is the average time of evaluation and $\Delta t_k$ is the time increment. The flux reconnected up to any time $t = \sum_{k=1}^{K}\Delta t_k$ is the sum 
\begin{equation}
\Psi_{rec}(t) = \sum_{k=1}^{K}{\dot{\Psi}(t_k)\Delta t_k}.
\end{equation}


\end{document}